\documentclass[%
 aip,
 jcp,
 reprint%
]{revtex4-1}

\usepackage{amsmath}
\usepackage{amssymb}
\usepackage{graphicx}% Include figure files
\usepackage{dcolumn}% Align table columns on decimal point
\usepackage{bm}% bold math
\usepackage[utf8]{inputenc}
\usepackage[T1]{fontenc}
\usepackage{mathptmx}
\usepackage{etoolbox}

\usepackage{siunitx}
\usepackage{braket}
\usepackage[version=4]{mhchem}

\usepackage{float}

\makeatletter
\def\@email#1#2{%
 \endgroup
 \patchcmd{\titleblock@produce}
  {\frontmatter@RRAPformat}
  {\frontmatter@RRAPformat{\produce@RRAP{*#1\href{mailto:#2}{#2}}}\frontmatter@RRAPformat}
  {}{}
}%
\makeatother

\begin{document}

\preprint{AIP/123-QED}

\title[Prediction challenge: First principles simulation of the ultrafast electron diffraction spectrum of cyclobutanone]{Prediction challenge: First principles simulation of the ultrafast electron diffraction spectrum of cyclobutanone}
% Force line breaks with \\
\author{Jiří Suchan}
\affiliation{%
Institute of Advanced Computational Science, Stony Brook University, Stony Brook, New York 11794, United States%\\This line break forced% with \\
}%

\author{Fangchun Liang}
\affiliation{%
Institute of Advanced Computational Science, Stony Brook University, Stony Brook, New York 11794, United States%\\This line break forced% with \\
}%
\affiliation{Department of Chemistry, 
Stony Brook University, Stony Brook, New York 11794, United States}

\author{Andrew S. Durden}
\altaffiliation[Current address: ]{Department of Chemistry, Wayne State University, Detroit, Michigan 48202, United States}
\affiliation{%
Institute of Advanced Computational Science, Stony Brook University, Stony Brook, New York 11794, United States%\\This line break forced% with \\
}%
\affiliation{Department of Chemistry, 
Stony Brook University, Stony Brook, New York 11794, United States}

\author{Benjamin G. Levine}
 %\homepage{http://www.Second.institution.edu/~Charlie.Author.}
\affiliation{%
Institute of Advanced Computational Science, Stony Brook University, Stony Brook, New York 11794, United States%\\This line break forced% with \\
}%
\affiliation{Department of Chemistry, 
Stony Brook University, Stony Brook, New York 11794, United States}
\email{ben.levine@stonybrook.edu}

\date{\today}% It is always \today, today,

\begin{abstract}
Computer simulation has long been an essential partner of ultrafast experiments, allowing the assignment of microscopic mechanistic detail to low-dimensional spectroscopic data.
However, the ability of theory to make {\em a priori} predictions of ultrafast experimental results is relatively untested.
Herein, as a part of a community challenge, we attempt to predict the signal of an upcoming ultrafast photochemical experiment using state-of-the-art theory in the context of preexisting experimental data. 
Specifically, we employ {\em ab initio} Ehrenfest with collapse to a block (TAB) mixed quantum-classic simulations to describe the real-time evolution of the electrons and nuclei of cyclobutanone following excitation to the 3s Rydberg state. The gas-phase ultrafast electron diffraction (GUED) signal is simulated for direct comparison to an upcoming experiment at the Stanford Linear Accelerator Laboratory.
Following initial ring-opening, dissociation via two distinct channels is observed: the C3 dissociation channel, producing cyclopropane and \ce{CO}, and the C2 channel, producing \ce{CH2CO} and \ce{C2H4}.
Direct calculations of the GUED signal indicate how the ring-opened intermediate, the C2 products, and the C3 products can be discriminated in the GUED signal.
We also report an {\em a priori} analysis of anticipated errors in our predictions: without knowledge of the experimental result, which features of the spectrum do we feel confident we have predicted correctly, and which might we have wrong?
\end{abstract}

\maketitle

\section{\label{sec:level1}Introduction}
Ultrafast spectroscopic measurements \cite{Zewail2000} ushered in a new era of chemistry, in which the microscopic motions of molecules could be observed directly.  But while one might dream of a ``molecular movie,'' in which the states of individual electrons and nuclei are recorded as a function of time in intricate detail, realistic ultrafast measurements necessarily involve the projection of molecular dynamics into a reduced-dimensional representation via a limited set of observables.  As such, from very early on, the simulation of ultrafast dynamics has been an invaluable partner of experiment.\cite{Janssen1993,Schwartz1994a,Batista1997,Vreven1997,BenNun1998}  Where experiment provides a lossy projection of real chemical dynamics, simulated dynamics may be analyzed in their full dimensionality.  But simulated dynamics are necessarily approximate in all but the smallest systems.  Despite three decades of intense effort, the inherent complexity of modeling many-body quantum dynamics drives an ongoing effort to develop practical approximations,\cite{Subotnik2016,Gossel2018,barbatti_review,basiletodd_rev,gonzales_book,Agostini2021} benchmark them for accuracy,\cite{Ibele2020,Zhang2021,Janos2023} and develop high-performance, user-friendly software implementations.\cite{Akimov2016,mai2018,fedorov2020,malone2020,Barbatti2022} Simulation has been very successfully applied as a tool for {\em a posteriori} analysis, connecting experimental signatures to actual molecular motions that cannot be unambiguously inferred from low-dimensional experimental data.\cite{Barbatti2010,Richter2012,Penfold2012,nelson2014,tavernelli2015,wang2015,Fielding2018,schuurman2018,popp2021,Talotta2022,dergachev2023}

Dynamics simulations alone are valuable, but the direct computation of ultrafast experimental observables from molecular dynamics simulations enables a much more definitive assignment of spectroscopic features.\cite{Hudock2007,Mitric2011,Rubio2013,Subotnik2014,Govind2015,Parkhill2016,Dsouza2018,Sanchez2018,Domcke2021,Kubas2021,Cerullo2021,Xu2022,Chakraborty2022,Silfies2023}  Without direct simulation of the experimental observable, the connection between theory and experiment is often made by comparing lifetimes.  Because lifetimes often depend exponentially on computed energies, which may themselves carry significant errors, lifetimes are among the most error-prone quantities one can compute.  Most other observables do not suffer from an exponential amplification of errors in the potential energy surface (PES), therefore, explicit simulation of experimental observables (e.g. time-resolved spectra or scattering signals) enables a more direct connection between theory and experiment.  Armed with computed observables, one may make definitive assignments even when simulated lifetimes have significant errors. 

As light sources continue to advance,\cite{young2018roadmap,nisoli2017attosecond} novel theoretical methods are required to connect to the new experiments they enable.\cite{kowalewski2017,Neville2018,realtimemethods,Zinchenko2021,Cheng2022,Freixas2022}  Gas phase ultrafast electron diffraction (GUED) experiments\cite{wolf_ued, Williamson1997,Ihee2001,Yang2016a, Yang2016b, Yang2018, Wolf2019} offer a particularly appealing target for direct simulation, given the intuitiveness of the observable (interatomic distances).  Because even very low levels of electronic structure theory can predict bond lengths with errors of only a few percent, the simulation of GUED enables an extraordinarily tight connection between experiment and theory.\cite{STEFANOU2017300,Wolf2019,parrish2019,champenois2021aims,liu2023rehybridization} 

With dynamical simulation now established as an essential partner to experiment in ultrafast science, it is natural to ask a more ambitious question: to what extent can simulation predict ultrafast dynamics {\em a priori}?  With this question in mind, the community conceived a prediction challenge to which this manuscript responds.  Over the last 15 weeks, we have attempted to predict the signal of an upcoming GUED experiment on cyclobutanone to be carried out at the Stanford Linear Accelerator Laboratory\cite{slac} (SLAC) in the days following the submission of this work.\footnote{This work was first submitted to arXiv on Jan. 15, 2024.  https://doi.org/10.48550/arXiv.2401.08069}  Herein we report the resulting prediction.  We also report a brief analysis of the predicted dynamics and how they will manifest in the experimental signal.  Finally, with the goal of testing our own understanding of our methods, we present predictions of what errors we might anticipate in our computed experimental signal.  By taking part in this collective effort, we hope to better understand the limits of ultrafast theory and identify directions for the development of practical tools capable of definitively predicting ultrafast dynamics.

\section{Methods}
\subsection{Nonadiabatic dynamics}
We will simulate the nonadiabatic dynamics of cyclobutanone using a novel {\em ab initio} implementation of the Ehrenfest with collapse to a block (TAB) method, which is described in detal in refs. \citenum{esch2020a} and \citenum{tab21}.  Here we provide a brief introduction to the basis features of the method.  We ran our simulations in an open source software package, TAB-DMS.\cite{tabdms_zenodo} %\footnote{Github repository: https://github.com/blevine37/tab-dms/tree/tab} 

The TAB mixed quantum-classical method was developed to correct errors associated with the mean-field nature of Ehrenfest dynamics\cite{tully_errors} via a decoherence correction that provides an accurate and efficient treatment of systems with many electronic states. Nuclei are propagated classically, with the force vector, $\bm{F}_{\rm{MF}}$, calculated in a mean-field fashion:
\begin{equation}
\label{eq1ehrenfest}
\bm{F}_{\rm{MF}}(t) = -\frac{\textrm{d}}{\textrm{d}\bm{R}}\frac{\bra{\Psi(t)}\bm{H}(t)\ket{\Psi(t)}}{\braket{\Psi(t)|\Psi(t)}}\,,
\end{equation}
averaging the effect of electronic Hamiltonian, $\bm{H}(\bm{R}(t))$, on the time-dependent electronic wave function $\Psi(t)$. Using the Verlet algorithm, we propagate the nuclear positions, $\bm{R}$, in discrete time steps. 

The TAB decoherence correction is based on an assumption of exponential decay of electronic coherences at a rate given by the difference between gradients of each pair of potential energy surfaces (PES).\cite{bittnerrossky1,bittnerrossky2} At the end of each nuclear time step, $\Delta t$, the loss of coherence is incorporated into an approximate electronic density matrix, $\boldsymbol{\rho}^{d}$, which represents the mixed electronic state that arises from decoherence during a single time step.  This matrix is generated from the coherent mean-field electronic density matrix, $\boldsymbol{\rho}^{c}$, associated with a given trajectory. The populations are taken directly from the coherently propagated mean-field density matrix:
\begin{equation}
\label{eq1tab}
\rho_{ii}^{d}=\rho_{ii}^{c}(t+\Delta t)\,,
\end{equation}
while the off-diagonal elements are scaled according to:
\begin{equation}
\label{eq2tab}
\rho_{ij}^{d}=\rho_{ij}^{c}(t+\Delta t)e^{\frac{-\Delta t}{\tau_{\mathrm{eff},ij}}}\,.
\end{equation}
The pivotal quantity here is the effective state-pairwise decoherence time, $\tau_{\mathrm{eff},ij}$.  Our approach starts from the widely-used estimate of the decoherence time derived by Bittner and Rossky\cite{bittnerrossky1,bittnerrossky2}:
\begin{equation}
\label{eq3tab}
\tau_{ij}^{-2}=\sum_{\eta}\frac{\left(F_{i,\eta}^{\mathrm{avg}}-F_{j,\eta}^{\mathrm{avg}}\right)^2}{\hbar^2\alpha_{\eta}}\,.
\end{equation}
The sum runs over all degrees of freedom, indexed $\eta$, adding the differences of electronic-state-specific forces, $\bm{F}^{\rm{avg}}_i$, averaged over the beginning and the end of a timestep. \textcolor{black}{Parameter $\alpha_{\eta}$ represents the atom-specific width of the Gaussian wavepacket and can therefore be determined without empirical fitting by fitting the harmonic ground state vibrational wave function.\cite{Esch2019}  In doing this fitting, we constrain the widths such that each degree of freedom associated with a particular atom type (C, H, O) has the same width as all others of the same type.}  The reduced Planck constant is $\hbar$. The decoherence time represents the decay of the overlap between Gaussian wave packets on a pair of linear, non-parallel PESs. This overlap is a Gaussian function of time, and the decoherence time corresponds to its width.  The effective decoherence time, $\tau_{\mathrm{eff},ij}$, is computed from the Bittner-Rossky decoherence time, $\tau_{ij}$, by integrating over an approximate history of the dynamics, as detailed in ref.\citenum{tab21}  In so doing, we are able to effectively treat the coherence decay as Gaussian, rather than exponential. 

Next, we use a stochastic procedure (described in detail in ref. \citenum{esch2020b}) to collapse the electronic wave function to a new pure linear combination of a subset of occupied adiabatic states. This procedure is designed such that the ensemble average of the resulting electronic density matrix (averaged over all possible outcomes) is $\boldsymbol{\rho}^{d}$. The velocity is rescaled at every time step where collapse occurs for conservation of the total energy.  Though we have recently developed an approximate scheme for extending TAB to very large numbers of electronic states,\cite{esch2020b} it is not necessary here.

In order to compute the GUED spectrum, a set of 150 trajectories was initiated on the S$_2$ (3s Rydberg) electronic state.  Initial nuclear positions and momenta were sampled from the Wigner distribution function computed in the harmonic approximation. The ground state minimum geometry and frequencies were computed via B3LYP with \mbox{6-311++G**} basis. No attempt was made to account more specifically for the excitation wavelength via the initial conditions.  All observables are taken as averages over this ensemble.  

\subsection{Electronic structure}
\subsubsection{Propagation algorithm}

To propagate the electronic structure, we use our high-performance implementation of the time-dependent complete active space configuration interaction method (TD-CASCI).\cite{tdci,durden2022} The electronic wave function, $\Psi(t)$, is represented as a linear combination of Slater determinants, $\Phi_{I}$, with time-dependent expansion coefficients $C_{I}(t)$:
\begin{equation}
\label{eq1tdci}
\Psi(t) = \sum_{I} C_{I}(t) \Phi_{I}\,.
\end{equation}
Their propagation is handled by numerically solving the time-dependent Schr{\"o}dinger equation:
\begin{equation}
\label{eq2tdci}
i\dot{\bm{C}}(t) =\bm{H}(t)\bm{C}(t)\,,
\end{equation}
via symplectic split operator integration.\cite{gray1996} 

The TAB dynamics driver is interfaced with the TD-CASCI propagation scheme, providing a robust and efficient method to simulate nonadiabatic dynamics. The TD-CASCI algorithm is implemented in a development version of the \mbox{TeraChem} electronic structure package.\cite{terachem21,Ufimtsev2008,Ufimtsev2009a,Ufimtsev2009b,Fales2015,Hohenstein2015,tdci,durden2022} The combination of GPU-acceleration and direct implementation of $\bm{HC}$ products allows efficient propagation in the full space of states, without the need to build any data structures with the full dimension of the electronic Hamiltonian.  

In theory, this combination allows us to model any incoming pulse directly in the dynamics while capturing possible initial coherences. Due to the time constraints of the project, all of our dynamics start directly on a given state. This is further justified by the known form of the experimental pulse, which intends to excite the first 3s Rydberg state. To ensure energy conservation, the electronic time step was set to 2\,as and the nuclear time step to 0.12\,fs. The simulations aim to capture \textcolor{black}{500 fs} of the post-excitation process.

\subsubsection{Electronic ansatz selection}
Cyclobutanone photochemistry has been examined in several past quantum chemical studies.\cite{prevaims,prevringop,prevnorish} Several important points on the PES have been identified: S$_0$ and S$_1$ minima and a transition state which seems to govern the overall pre-dissociative dynamics on S$_1$. \textcolor{black}{Visualization of the main reaction pathways is depicted on Figure 2.  Our simulations predict that most of the S$_2$ population transitioning to the S$_1$ state via a closed-ring conical intersection. In S$_1$, the molecule must overcome a ring-opening barrier before dissociating via two main reaction channels labeled as C2 for CH$_2$CO + C$_2$H$_4$ products and C3 for C$_3$H$_6$ (cyclopropane) + CO products.} Based on the past \textcolor{black}{publications}, it appears that multi-state second-order perturbation theory (MS-CASPT2) on the complete active space self-consistent field (CASSCF) wave function accurately captures the significant correlation effects. Quite large active spaces, ranging from 6 electrons in 6 orbitals (6,6) up to 12 electrons in 11 orbitals (12,11), were used to capture the bond-breaking reaction channels. These studies usually state-averaged only over two valence singlet electronic states, thus less is known about the 3s Rydberg state. 

Regarding the role of triplet states, intersystem crossing from the S$_1$ state seems to take place on nanosecond timescales.\cite{prevnorish,isctime} It dominates only near the onset of the absorption spectrum where there is not enough energy to traverse the S$_1$ barrier and relax via internal conversion (IC).  Therefore, we will not consider ISC in our simulations.

\begin{figure}
\includegraphics[width=0.45\textwidth]{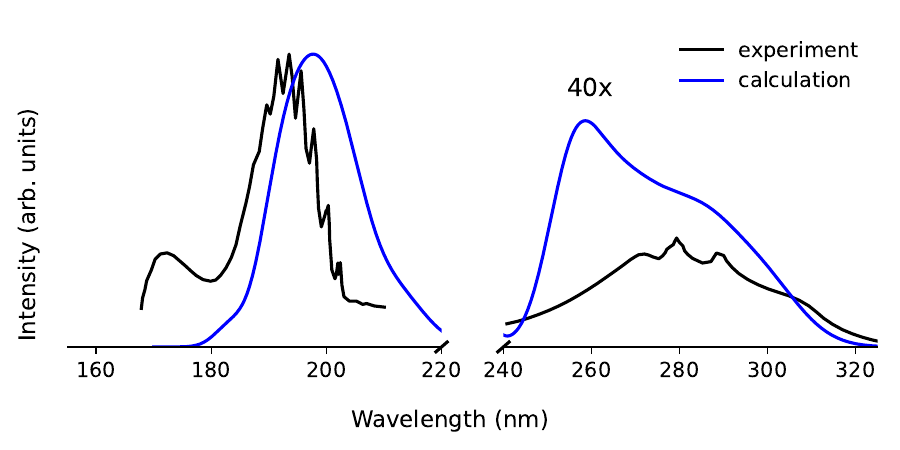}%
\caption{Normalized experimental\cite{elsayed_ryd,leeval} and calculated absorption spectrum using the selected FOMO-CASCI(8,11)/6-311+G* method.}%
\label{fig1:abspec}
\end{figure}

The upcoming experiment will excite to the 3s Rydberg state (S$_2$) with 200\,nm (6.2 eV) laser light, which is well above the darker $n\rightarrow\pi^*$ (S$_1$) state (280\,nm; 4.4 eV). The experimental absorption spectrum\cite{elsayed_ryd,leeval} (Fig.\ref{fig1:abspec}) reveals that the selected wavelength corresponds to the onset of the 3s Rydberg state. We further observe the 3p Rydberg states in the range of 170-180\,nm (6.9--7.3 eV). Since states seem well separated, we need to choose an electronic structure ansatz that will reliably describe the overall $\rm{S}_2\rightarrow\rm{S}_1\rightarrow\rm{S}_0$ internal conversion process, but need not describe the 3p Rydberg states and above.  

 We tested several combinations of multireference methods and active spaces, as the choice of orbitals can play an important role in determining the accuracy of CASCI-based dynamics simulations.\cite{Levine2021}  Since the most important task is to capture the pre-dissociation phase accurately, we aimed for the best agreement of relative state energies in several structures. To this end, we choose a floating occupation molecular orbitals complete active space configuration interaction method (FOMO-CASCI) ansatz.\cite{fomo,granucci2001} We choose an active space of 8 electrons in 11 orbitals (8,11), including the ground state-occupied non-bonding orbitals of the O atom and the $\pi$ and $\sigma$ bonding orbitals on the carbonyl group. The ground state-virtual orbitals in the active space include the carbonyl $\pi^*$ orbital and two $\sigma^*$ orbitals, as well as the 3s, 3p$_x$, 3p$_y$, 3p$_z$ Rydberg orbitals. This selection is capable of describing the relevant Rydberg and valence states, as well as the bond dissociation processes to expected products. The FOMO-CASCI method uses a fictitious electronic temperature, which we set to 0.3\,Ha. The employed 6-311+G* basis set includes diffuse functions that are necessary for the description of Rydberg states.

To establish the quality of the selected FOMO-CASCI method, we make a comparison to the experiment.\cite{elsayed_ryd,leeval} Figure \ref{fig1:abspec} displays the calculated absorption spectrum, which is computed by compounding vertical spectra computed at a set of Wigner-sampled geometries with 0.1\,eV Gaussian line broadening. The excitation energies at the ground state minimum geometry are presented in Table \ref{tab:ene} with experimental absorption maxima for comparison.  In both cases, we observe that the FOMO-CASCI gap between valence and Rydberg states is slightly smaller than observed experimentally, but overall the FOMO-CASCI method is very accurate in the Franck-Condon region. 

\begin{table}[h]
\caption{\label{tab:ene} FOMO-CASCI vertical excitation energies calculated with the (8,11) active space and 6-311+G* basis at the B3LYP-optimized S$_0$ minimum geometry. %All calculations in TeraChem except CASPT2 in Open Molcas. 
Experimental data based on spectral maxima.}
%3st
%\begin{tabular}{c c c c c c}
%  & CASCI & FOMO-CASCI & CASSCF & CASPT2 & expt.\cite{elsayed_ryd,leeval} \\ 
% $n \rightarrow \pi^*$ & 8.13 & 4.65 & 4.26 & 4.24 & 4.4\\  
% $n \rightarrow 3$s & 8.57 & 6.32 & 6.89 & 6.79 & 6.2   \\
%% $\pi \rightarrow 3$p & & & 
%\end{tabular}
%\end{table}

%6st
%\begin{tabular}{c c c c c c}
%  & CASCI & FOMO & CASSCF & CASPT2 & expt.\cite{elsayed_ryd,leeval} \\ 
% $n \rightarrow \pi^*$ & 8.13 & 4.64 & 4.64 & 4.31 & 4.4\\  
% $n \rightarrow 3$s & 8.57 & 6.32 & 6.08& 6.73 & 6.2   \\
% $n \rightarrow 3$p & 9.26,9.39,9.41 & 7.06,7.15,7.16 & 6.81,6.92,7.01 & 7.34,7.62,7.62& 7.1 \\ 
%\end{tabular}
%\end{table}

%3st reduced
\setlength{\tabcolsep}{8pt}
\begin{tabular}{c c c c }
%\hline
% & FOMO-CASCI  & experiment\cite{elsayed_ryd,leeval} \\ 
\textcolor{black}{State} & \textcolor{black}{Character} & FOMO-CASCI  & Experiment\cite{elsayed_ryd,leeval} \\ 
 \hline
\textcolor{black}{S$_1$}& $n \rightarrow \pi^*$  & 4.65 &  4.4\\  
\textcolor{black}{S$_2$}& $n \rightarrow 3$s  &  6.32  & 6.2   \\
 \hline
%% $\pi \rightarrow 3$p & & & 
\end{tabular}
\end{table}

\begin{figure*}
%\begin{figure}
\includegraphics[width=1\textwidth]{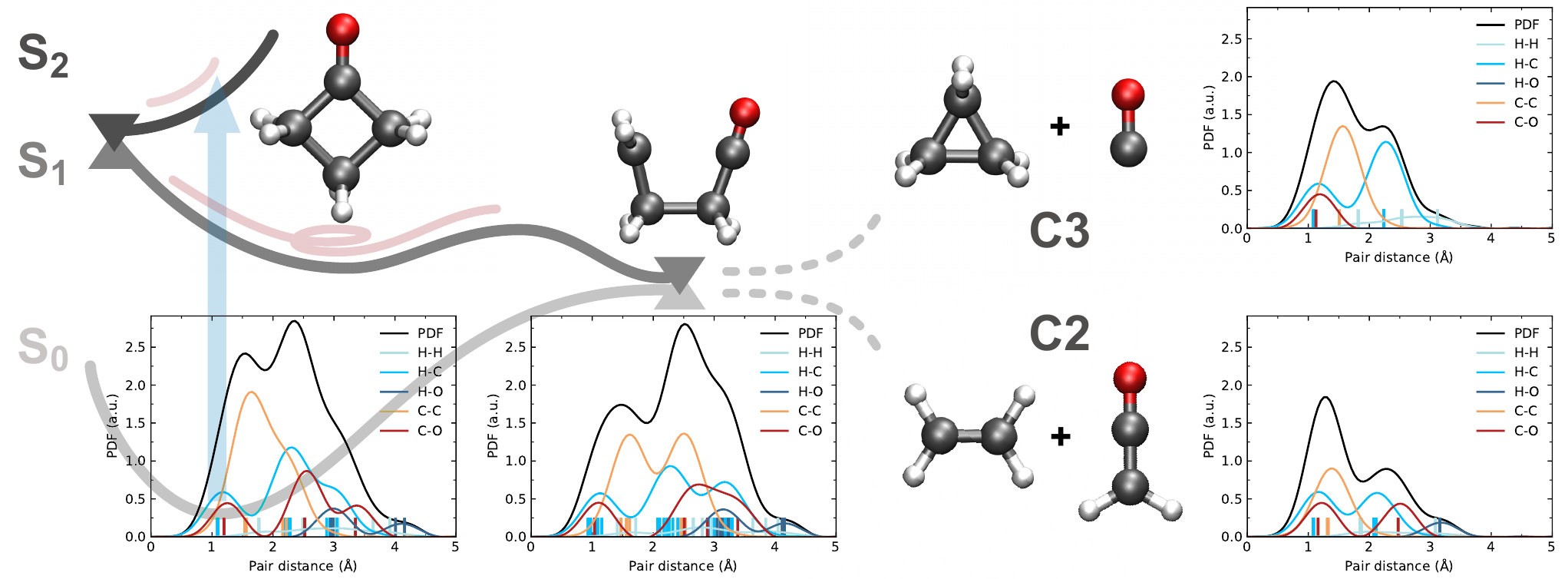}%
\caption{Visualization of the main photochemical pathways of cyclobutanone following excitation to S$_2$. Expected PDF signal calculated using IAM model for the initial molecule, open ring structure and products (C2, C3) with individual contributions from atomic pairs.}%
\label{fig:vis}
%\end{figure}
\end{figure*}

\subsection{Calculation of GUED signal}
To simulate the GUED signal, we calculate time-dependent 1D electron diffraction patterns. They can also be converted to pair distribution functions (PDFs). Both representations are directly comparable to the experiment. 

%To calculate the GUED signal, we utilize the independent atom model (IAM), closely following the procedure described by Centurion et al.\cite{wolf_ued} First, we calculate the electron scattering amplitudes from the individual atoms. These atomic form factors (AFFs), $f(s)$, were calculated using the ELSEPA program\cite{elsepa} using parameters for the SLAC facility. To simulate the scattering intensity, $I(s)$, of a molecule as a function of momentum transfer, $s$, we consider individual ($I_{\rm{at}}$) and pairwise ($I_{\rm{mol}}$) contributions of scattering amplitudes. For a gas phase sample we use the simplified equation:
%\begin{equation}
%\label{eq1}
%I(s) = I_{\rm{at}} + I_{\rm{mol}} = \sum_{i=1}^{N}\left|{f_i(s)}\right|^2 + \sum_{i=1}^{N}\sum_{j=1,j\neq i}^{N}f_i^*(s)f_j(s)\frac{\mathrm{sin}(sr_{ij})}{sr_{ij}}\,.
%\end{equation}
%Here, the $i,j$ indices loop over $N$ atoms. Interatomic distances are denoted as $r_{ij}$. Due to the nominal $s^{-5}$ dependence of the scattering intensity, we work with the modified scattering intensity,
To calculate the GUED signal, we utilize the independent atom model (IAM), closely following the procedure described by Centurion et al.\cite{wolf_ued} First, we calculate the electron scattering amplitudes from the individual atoms. These atomic form factors (AFFs), $f(s)$, were calculated using the ELSEPA program\cite{elsepa} using parameters for the SLAC facility. \textcolor{black}{The scattering intensity of a molecule, $I(s)$,} as a function of momentum transfer, $s$, \textcolor{black}{is composed of} individual ($I_{\rm{at}}$) and pairwise ($I_{\rm{mol}}$) contributions of scattering amplitudes. For a gas phase sample we \textcolor{black}{assume} the simplified equation:
\begin{equation}
\label{eq1}
I(s) = I_{\rm{at}} + I_{\rm{mol}} = \sum_{i=1}^{N}\left|{f_i(s)}\right|^2 + \sum_{i=1}^{N}\sum_{j=1,j\neq i}^{N}f_i^*(s)f_j(s)\frac{\mathrm{sin}(sr_{ij})}{sr_{ij}}\,.
\end{equation}
Here, the $i,j$ indices loop over $N$ atoms. Interatomic distances are denoted as $r_{ij}$. Due to the nominal $s^{-5}$ dependence of the \textcolor{black}{molecular} scattering intensity, we work \textcolor{black}{with a more suitable form}, the modified scattering intensity,
\begin{equation}
\label{eq2}
sM(s)=s\frac{I_{\rm{mol}}}{I_{\rm{at}}}\,.
\end{equation}
Finally, the real-space PDF, $P(r)$, which describes the probability of atom pairs being at a given distance, can be obtained by sine transformation,
\begin{equation}
\label{eq3}
P(r)=r\int_{0}^{\infty}sM(s)\textrm{sin}(sr)\textrm{d}s\,.
\end{equation}
To consider experimental limitations, we should integrate this equation between the limits of the detector, $s_{\rm{min}}$ and $s_{\rm{max}}$, and add a damping factor that avoids artifacts of the transformation,
\begin{equation}
\label{eq4}
P(r)=r\int_{s_{\textrm{min}}}^{s_{\textrm{max}}}sM(s)\textrm{sin}(sr)\textrm{e}^{-\alpha s^2}\text{d}s\,.
\end{equation}
The experimental resolution for the upcoming measurement is $1-10\,\si{\angstrom}^{-1}$. Nevertheless, the data between 0 and $s_{\rm{min}}$ are usually interpolated to avoid artifacts in the PDF. Therefore, in our calculations we used the full $0-10\,\si{\angstrom}^{-1}$ range to emulate the instrumental limitations, and we employ a damping factor $\alpha$ of 0.04\,$\si{\angstrom}^{2}$. \textcolor{black}{PDFs computed at specific important geometries are depicted on Figure \ref{fig:vis}.} \textcolor{black}{In dynamics simulations, t}he modified scattering intensity is calculated as an average over the ensemble of trajectories at a given time. We also calculate the difference pair distribution function ($\Delta$PDF) by subtracting the scattering signal at time zero (in Eq. \ref{eq2}) to better observe various relaxation channels. We report two different representations of the experimental signal: the signal computed with the temporal resolution of the simulation (assuming the pump pulse instantaneously excites the molecule to S$_2$), and a second version convoluted with a 80\,fs FWHM temporal Gaussian, matching the anticipated experimental cross-correlation. %Visualization of expected reaction pathways with PDFs for reactant and products is depicted on Figure \ref{fig:vis}. 

\section{Results}
\subsection{TAB simulations}
\begin{figure}[h]
\includegraphics[width=0.43\textwidth]{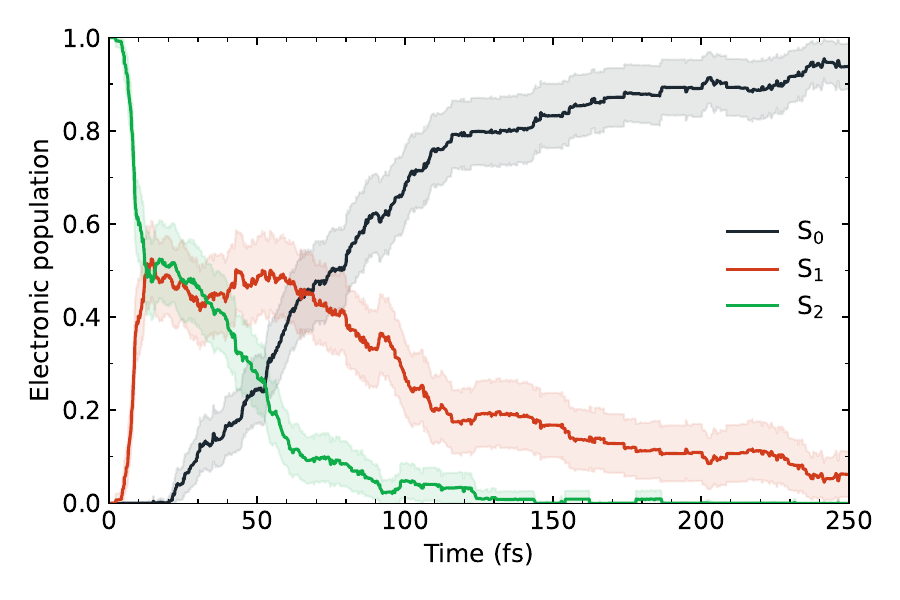}
\caption{Electronic population evolution during TAB dynamics initiated in the S$_2$ state.}
\label{fig:s2pop}
\end{figure}

\begin{figure*}
\includegraphics[width=.95\textwidth]{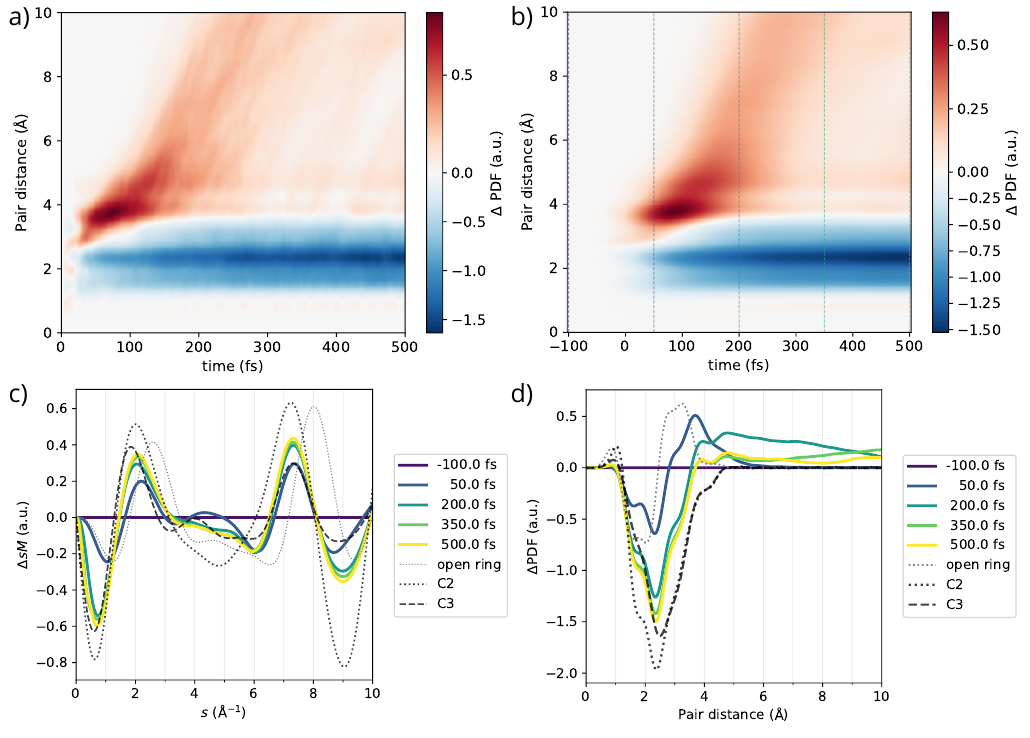}
\caption{a) $\Delta$PDF simulated signal. b) Same signal convoluted with a 80\,fs FWHM temporal Gaussian. \textcolor{black}{Color-coded vertical lines correspond to the time slices plotted in panels c and d.}  c)~Time slices of \textcolor{black}{convoluted} modified scattering intensity from \num{-100}\,fs to 500\,fs. Reference $\Delta sM$ signals for optimized C2 products, C3 products, and open ring structures in black. d) Time slices of $\Delta$PDF presented in the same manner.}%
\label{fig:tdpdf}
\end{figure*}

The average adiabatic electronic populations predicted by our TAB simulations are displayed in Fig. \ref{fig:s2pop}. The lines represent the mean values over the ensemble of trajectories, and the shaded areas show the 95\,\% confidence interval based on the binomial distribution. (This confidence interval reflects only error associated with trajectory sampling, not error related to physical approximations.)  Following excitation, the molecule undergoes rapid relaxation through a conical intersection (CI) to the S$_1$ state. Half of the original population transitions during the first 12\,fs. However, it takes 65\,fs until the S$_2$ population decays below 0.1. \textcolor{black}{Analysis of the minimum energy conical intersection between S$_2$ and S$_1$ (Figure S1) indicates that it is strongly symmetry broken, with one of the two symmetry-equivalent CO-CH$_2$ bonds stretched to 1.83 \AA.  This indicates that there is an initial impulse toward ring opening on S$_2$.} Once on S$_1$, the molecule relaxes to the ground state via the respective CI. This requires crossing of a ring-opening transition state, which may prolongs the relaxation time. After reaching the ground state, we observe the usual relaxation channels \textcolor{black}{to} C2 \textcolor{black}{and} C3 \textcolor{black}{products}.%labeled as C2 for \ce{CH2CO} + \ce{C2H4} products and C3 for \ce{C3H6} (cyclopropane) + \ce{CO} products.  

In addition to these main pathways, 19\% of the population is observed to undergo ring-opening directly in the S$_2$ state, which contributes to the prolonged S$_2$ relaxation. The molecule then transitions to either S$_1$ or S$_0$ at various points in time.  Both C2 and C3 channels are observed following relaxation to S$_0$.  Within our sampling of 150 trajectories, it is not obvious that ring-opening on S$_2$ impacts the photoproduct distribution.

In total, most trajectories form the C3 products (60\%), although the cyclization to cyclopropane is not always completed within our \textcolor{black}{500 fs} simulation window. C2 products are created in 28\% of cases. In the remaining 12\% we observe some larger fragmentation of the original molecule \textcolor{black}{including CO formation and} hydrogen dissociation. \textcolor{black}{The presented product analysis does not include trajectories that terminated prematurely or did not lead to the complete ring-opening reaction. Ultimately, 70 and 32 trajectories followed the C3/C2 channels, respectively, yielding a branching ratio of 2.2.} 

\textcolor{black}{We note that our trajectories were facing convergence issues. Only 32\% of them reached the 500 fs mark (see Figure S2 for details). Cyclobutanone photodissociation seems to present a challenge for electronic structure methods mainly due to the formation of various molecular species in the process. This additional source of uncertainty in our prediction is the direct result of the artificial time constraints of the prediction challenge.}

From the TAB dynamics, we simulate the time-dependent $\Delta$PDF signal. Fig. \ref{fig:tdpdf} displays the resulting signal at simulation resolution (panel a) and convoluted by the experimental cross-correlation (panel b). There is a faint signal in the first tents of fs when the initial S$_2\rightarrow$S$_1$ relaxation takes place. After that, we start to observe a new signal above the previously blank 3\,\si{\angstrom} threshold. This is due to the ring-opening process and subsequent dissociation into distinct molecular products. 
Since the $\Delta$PDF fingerprints of individual intermediates and products are rather similar, it is difficult to discern between them using this plot. For easier analysis, we present time slices of the $\Delta$PDF data in Fig. \ref{fig:tdpdf}d. The corresponding plot in momentum space---the difference modified scattering intensity ($\Delta sM$)---is shown in Fig. \ref{fig:tdpdf}c. Reference signals associated with three static geometries are presented for comparison: a representative open ring intermediate structure and the C2 and C3 products.  We have chosen the FOMO-CASCI-optimized S$_1$-S$_0$ minimal energy conical intersection structure (pictured in Fig. \ref{fig:vis}) to represent the ring-opened region of the PES. We do not mean to imply that signal in this region necessarily corresponds to detection of population at the conical intersection. However, it carries the ring-opening character which is specific for the product formation. The rest of the reference structures were optimized at B3LYP/\mbox{6-311+G*} level.

At 50\,fs, both the $\Delta$PDF and difference modified scattering intensity ($\Delta sM$) exhibit features that are characteristic of the representative ring-opened (S$_1$-S$_0$ conical intersection) structure.  In momentum space, the first negative peak (between 0 and 2 \si{\angstrom}$^{-1}$) closely mirrors that of the ring-opened structure, and there is a pronounced negative peak at near 6 \si{\angstrom}$^{-1}$ that does not appear in either the C2 or C3 products.  In the $\Delta$PDF, there is a large positive peak just below 4 \si{\angstrom} that is characteristic of the ring-opened structure.

At later times, the signals evolve in a way that reflects the dissociation. In momentum space, the first negative peak intensifies and shifts to lower momentum (now below 1 \si{\angstrom}$^{-1}$) upon dissociation.  In position space, the large peak near 4 \si{\angstrom} loses intensity as the ring-opened structure dissociates.  In addition, the broad negative feature between 1 and 3 \si{\angstrom} intensifies.

A detailed analysis of this broad negative feature allows discrimination between the C2 and C3 products.  The lower-distance portion of this feature ($\sim$1.5 \si{\angstrom}) corresponds to the loss of carbon-carbon bonds.  The reactants contain four C-C bonds.  The C3 products have three C-C bonds (all in cyclopropane), while the C2 product has only two (one in each product).  
Thus, as the ring-opened structure disappears, a more pronounced shoulder at $\sim$1.5 \si{\angstrom} is indicative of a lower C3/C2 branching ratio.  Differences are also observed in the momentum-space representation.  Between 4 and 5 \si{\angstrom}$^{-1}$, both signals are negative, but the C2 products contribute a larger negative signal.  Differences in the height and location of the first negative and positive peaks (near 0.5 and 2 \si{\angstrom}$^{-1}$, respectively) are also observed when comparing the C2 and C3 products. 

The presented results account for the first 500 fs of post-excitation dynamics, which could be reliably described using our limited number of trajectories. For longer times, we can expect slow decay of the dissociative signal (above 4 \si{\angstrom} in the $\Delta$PDF). Otherwise, the overall shape of the 500\,fs time slices should remain relatively unchanged.

\subsection{Discussion}
\label{subsec:discussion}

The primary goal of the present work is to test our ability to predict an experimental measurement, {\em a priori}.  However, it is worth discussing what kinds of errors we anticipate might arise from our approach before we have access to the result.  Just as it is interesting to see if we can predict the experimental signal, it is interesting to see if we can predict which errors might manifest in that prediction.  By recording our thinking, this prediction challenge will teach us not only about the accuracy of our methods, but also how well we understand them.

Let us first consider the reaction mechanism. A known advantage of \textit{ab initio} molecular dynamics is that it allows us to simulate molecular dynamics with relatively little prior knowledge of the relevant reaction pathway(s) and without enduring a time- and resource-consuming PES fitting procedure. In the present case, we predict that the molecule undergoes ring-opening, followed by dissociation via the C2 and C3 channels. The final photochemical outcomes are not truly predictions, because they are known from past photochemical experiments,\cite{denschlag1968benzene,cycbut_productratio_exp,laserphotolee} but the initial ring opening step is a prediction, and would likely be observable in the experiment.  Given that our simulations predict the initial and final states correctly, it seems likely that the prediction of the intermediate ring-opened intermediate state is correct as well.  Thus, we expect this prediction to be robust.

Now, we consider the accuracy of our prediction of the experimental observable itself.  As mentioned above, GUED is a powerful experiment in part because it can very accurately measure a time-dependent property (interatomic distance) that can also be very accurately predicted by simulation.  \textcolor{black}{Given a particular molecular structure, we are confident that the computed experimental observables ($\Delta$PDF and its momentum-space counterpart) are accurate.  This is not to say that we necessarily predict the correct structures at the correct time.}

Though we feel fairly confident about our predictions of mechanism and observable, our simulations also predict more challenging quantities: branching ratios and timescales.  We expect these predictions to be challenging because the calculation of these properties involves the effective exponentiation of relative energies between points on the PES, via $\mathrm{e}^{-\Delta E/k_{\mathrm{B}}T}$ ($\Delta$E denoting relative energy, $k_{\rm{B}}$ the Boltzmann constant, and $T$ temperature). This exponentiation can amplify small errors in the PES into large errors in the property of interest.  Our simulations are based on a relatively low level of electronic structure theory and some errors in the PES are inevitable. Therefore, predictions of rates and branching ratios are potentially less robust. A notable exception to this statement would be when the process of interest is governed by ballistic, rather than statistical, motion. The timescale of a ballistic process is governed by the gradient of the PES and the masses of the atoms. In such cases, there is no exponential amplification of error, and a robust prediction is more likely. In turn, we will consider potential errors in our predictions of the C3/C2 branching ratio, the timescale for $\rm{S}_2\rightarrow\rm{S}_1$ relaxation, and the timescale for subsequent $\rm{S}_1\rightarrow\rm{S}_0$ relaxation.

Both experiment and past theory suggest that the C3/C2 product ratio has a strong and complicated dependence on pump wavelength.\cite{laserphotolee,denschlag1968benzene, cycbut_productratio_exp}  This behavior can be attributed to the nonequilibrium vibrational energy distribution driving the dynamics in short time.\cite{prevaims,prevnorish} 
 Our simulations do not take the pump wavelength into account, except rather coarsely by the selection of our initial electronic state.  Thus, we anticipate that error in our predicted branching ratio is a likely source of error in the predicted spectrum.

Next, we will consider the $\rm{S}_2\rightarrow\rm{S}_1$ relaxation process. To this end, we can estimate the excited state lifetime from existing spectroscopic data.\cite{elsayed_ryd,leeval} Heller describes the relation between decay times of the wavepacket's autocorrelation function and the width of absorption spectrum peaks.\cite{heller1981semiclassical} Since there is a clear vibronic resolution of the first Rydberg state, we can infer \textcolor{black}{a lower bound on} the state's half-life in this way. The full width at the half maximum (FWHM) of a single subpeak can be estimated to around 0.02\,eV. We can infer the state's half-life by taking the reciprocal of FWHM, yielding 34\,fs. Though the decay of S$_2$ is non-exponential in our simulations, the simulated dynamics are consistent with the inferred experimental timescale; the S$_2$ population decays to 0.5 in the first 12\,fs, and then to 0.25 by $\sim$50 fs.  Given that this relaxation occurs in the first 1--3 vibrational periods, it appears to be a ballistic, barrierless process; thus it is not surprising that our predication here appears robust.  \textcolor{black}{However, a previous TRPES experiment by Kuhlman, et al. suggests the possibility of a longer sub-ps lifetime in the range 310-740 fs.\cite{Kuhlman2012}}

Finally, we consider the $\rm{S}_1\rightarrow\rm{S}_0$ relaxation process. In the case of cyclobutanone the height of the S$_1$ transition state and the available vibrational energy will determine the overall kinetics of the process. 
To estimate possible errors in our S$_1$ decay rate, we assume the rate-limiting step is traversing the S$_1$ transition state and apply Rice–Ramsperger–Kassel–Marcus (RRKM) theory\cite{rrkm} to predict the lifetime on a more accurate PES as a function of excess pump energy.  \textcolor{black}{This can be thought of as a sanity test on our results; if RRKM on a more accurate surface predicts a much slower (multi-ps) lifetime, this would be a strong indication that our dynamic simulations are much too fast.}  RRKM assumes the internal energy is randomly distributed among the vibrational modes of the molecule, and the reaction occurs when enough energy accumulates in the reactive mode. The rate constant is evaluated using the formula
\begin{equation}
\label{eqrrk}
%k(E)=\frac{\sum_{i=1}^s\nu_i}{\sum_{i=1}^{s-1}\nu_i^{\ddag}}\left(1-\frac{E_0}{E}\right)^{s-1}\,,
k(E)=\frac{\sigma W^{\ddag}(E-E_0)}{h\rho(E)},
\end{equation}
where $W^{\ddag}$ denotes the number of accessible vibrational states of the transition state (S$_1$TS), excluding those leading to the reaction, and $\rho$ the density of states of the reactant (S$_1$min). The available internal energy of the system is $E$, and $E_0$ is the barrier height, both relative to the S$_1$ minimum in our case. Planck's constant is $h$, and $\sigma$ represents degeneracy of the reaction pathway. We set $\sigma$ to 2, since there are two symmetry-equivalent ring-opening reactions.
Here, we use previously optimized structures from ref. \citenum{prevringop}, recalculating the energies and frequencies with the highly accurate complete active space second-order perturbation theory (CASPT2) method. We choose an active space of 10 electrons in 8 orbitals (CASPT2(10,8)) and the 6-31G* basis. Zero-point energy is included. To calculate the number of states, we use the Beyer-Swinehart algorithm.\cite{bsa} 

To calibrate our surface, we first estimate the half-life upon excitation at 281\,nm (4.41\,eV) and 255\,nm (4.85\,eV), which correspond to transient absorption measurements by Kao, et al.\cite{prevTAS} The computed half-lives (\textcolor{black}{1560} and \textcolor{black}{530}\,fs at 281 and 255\,nm, respectively) are within a factor of \textcolor{black}{four} of the shortest experimental timescale (450 fs, which was determined in a global fit including both experiments at 281\,nm and 255\,nm\cite{prevTAS}).  Given that one would expect a fairly strong wavelength dependence for this rate that is not discerned by the global experimental fit, it is difficult to say anything more specific than that RRKM/CASPT2 provides a reasonable estimate of the lifetime.

Now we consider the current simulations.  We assume that all excess energy in the 200\,nm (6.2\,eV) pump pulse is thermalized on S$_1$ prior to the reaction.  Within these assumptions, RRKM predicts an S$_1$ state half-life of \textcolor{black}{150}\,fs, which is longer than our simulated timescale by a factor of 2. One can attribute this discrepancy to two different factors: FOMO-CASCI may predict a barrier on S$_1$ that is too small, and/or the assumption of thermalization on S$_1$ in RRKM may be invalid here.  If all error is attributable to the assumption of thermalization, we would anticipate very little error in our predicted lifetime.  However, if the error is attributed to the FOMO-CASCI barrier height, it is imaginable that we underestimate the timescale for the $\rm{S}_1\rightarrow\rm{S}_0$ relaxation by a factor of 2, or even more. Keeping in mind that the timescales are exponentially sensitive to errors in the PES, this degree of uncertainty is not surprising.  
\textcolor{black}{Ultimately, these RRKM/CASPT2 results confirm the sanity of our prediction of a sub-ps lifetime.}
With additional time and computational effort, simulations based on CASPT2 or another highly accurate {\em ab initio} method would allow a more definitive prediction of the timescale.  

\section{Conclusions}
Above, we report a prediction of the GUED spectrum of cyclobutanone, excited to the 3s Rydberg state, for comparison to an upcoming experiment.  To this end, we applied an {\em ab initio} molecular dynamics approach based on the TAB nonadiabatic mixed quantum-classical method to simulate the dynamics following the initial pump pulse, and then we directly simulated the experimental observable.  Our simulations show two distinct dissociation pathways that share a common ring-opened intermediate.  Analysis of the predicted GUED signal provides guidance on how the ring-opened intermediate, C2 products, and C3 products can be discerned in the spectrum.

We also present our own {\em a priori} assessment of the most likely sources of error in our prediction.  Although the details are specific to our group's prediction, the general conclusions of our analysis apply broadly to all prediction attempts; the two most likely modes of failure for this prediction challenge are a) an incorrect prediction of the branching ratio between the C2 and C3 channels, and b) an incorrect prediction of the lifetime for barrier crossing on S$_1$.  This is because these two processes are governed by statistical motion, and therefore small errors in relative energies will be amplified via the Boltzmann factor $\mathrm{e}^{-\Delta E/k_\mathrm{B}T}$ into large errors in the quantity of interest.  \textcolor{black}{In short, we are hopeful that the procedure that we followed in this work is capable of providing a quantitative prediction of the experimental signal, except for those features of the signal that depend exponentially on energy (branching ratios and lifetimes).  Of course, significant errors in branching ratios and lifetimes may result in large changes to the signal.} Ultimately, these errors arise from the approximate electronic structure methods used in dynamic simulations, more so than the chosen dynamical method.  

Importantly, robust predictions of mechanisms and their observable signatures are likely to be sufficient to definitively assign the experimental spectrum, even in the case that predicted lifetimes and branching ratios have significant errors.  Thus, even a simulation that fails as a standalone prediction may provide essential insight into the dynamics of cyclobutanone when interpreted in the context of the experimental data.

\section{Supplementary Material}
The supplementary materials for this work include 1) a document containing optimized molecular structures, discussion of convergence issues experienced during dynamics, and details of the RRKM calculations; and 2) a zip file containing numpy data files with the predicted spectra.

\begin{acknowledgments}
The authors gratefully acknowledge support from the National Science Foundation under grant CHE-1954519 and the Institute for Advanced Computational Science (IACS). The development of the {\em ab initio} interface to TAB was supported by the U.S. Department of Energy, Office of Science, Office of Basic Energy Sciences, under Award No. DE-SC0021643.  J.S. acknowledges a postdoctoral fellowship from IACS.
\end{acknowledgments}

\section*{Author declarations}
\subsection*{Conflict of interest}
\noindent
The authors have no conflicts to disclose.
\subsection*{Author Contributions}
\noindent
\textbf{Jiří Suchan}: Investigation (lead); Software (supporting); Writing – original draft (equal); Writing – review \& editing (equal). \textbf{Fangchun Liang}: Software (equal). \textbf{Andrew S. Durden}: Software (equal). \textbf{Benjamin G. Levine}: Writing – original draft (equal); Writing – review \& editing (equal).

\section*{Data Availability Statement}

The data that support the findings of this study are available from the corresponding author upon reasonable request.

\appendix

\nocite{*}
\bibliography{citations}% Produces the bibliography via BibTeX.

\begin{thebibliography}{106}%
\makeatletter
\providecommand \@ifxundefined [1]{%
 \@ifx{#1\undefined}
}%
\providecommand \@ifnum [1]{%
 \ifnum #1\expandafter \@firstoftwo
 \else \expandafter \@secondoftwo
 \fi
}%
\providecommand \@ifx [1]{%
 \ifx #1\expandafter \@firstoftwo
 \else \expandafter \@secondoftwo
 \fi
}%
\providecommand \natexlab [1]{#1}%
\providecommand \enquote  [1]{``#1''}%
\providecommand \bibnamefont  [1]{#1}%
\providecommand \bibfnamefont [1]{#1}%
\providecommand \citenamefont [1]{#1}%
\providecommand \href@noop [0]{\@secondoftwo}%
\providecommand \href [0]{\begingroup \@sanitize@url \@href}%
\providecommand \@href[1]{\@@startlink{#1}\@@href}%
\providecommand \@@href[1]{\endgroup#1\@@endlink}%
\providecommand \@sanitize@url [0]{\catcode `\\12\catcode `\$12\catcode `\&12\catcode `\#12\catcode `\^12\catcode `\_12\catcode `\%12\relax}%
\providecommand \@@startlink[1]{}%
\providecommand \@@endlink[0]{}%
\providecommand \url  [0]{\begingroup\@sanitize@url \@url }%
\providecommand \@url [1]{\endgroup\@href {#1}{\urlprefix }}%
\providecommand \urlprefix  [0]{URL }%
\providecommand \Eprint [0]{\href }%
\providecommand \doibase [0]{http://dx.doi.org/}%
\providecommand \selectlanguage [0]{\@gobble}%
\providecommand \bibinfo  [0]{\@secondoftwo}%
\providecommand \bibfield  [0]{\@secondoftwo}%
\providecommand \translation [1]{[#1]}%
\providecommand \BibitemOpen [0]{}%
\providecommand \bibitemStop [0]{}%
\providecommand \bibitemNoStop [0]{.\EOS\space}%
\providecommand \EOS [0]{\spacefactor3000\relax}%
\providecommand \BibitemShut  [1]{\csname bibitem#1\endcsname}%
\let\auto@bib@innerbib\@empty
%</preamble>
\bibitem [{\citenamefont {Zewail}(2000)}]{Zewail2000}%
  \BibitemOpen
  \bibfield  {author} {\bibinfo {author} {\bibfnamefont {A.}~\bibnamefont {Zewail}},\ }\href {\doibase 10.1021/jp001460h} {\bibfield  {journal} {\bibinfo  {journal} {The Journal of Physical Chemistry A}\ }\textbf {\bibinfo {volume} {104}},\ \bibinfo {pages} {5660} (\bibinfo {year} {2000})}\BibitemShut {NoStop}%
\bibitem [{\citenamefont {Janssen}\ \emph {et~al.}(1993)\citenamefont {Janssen}, \citenamefont {Dantus}, \citenamefont {guo},\ and\ \citenamefont {Zewail}}]{Janssen1993}%
  \BibitemOpen
  \bibfield  {author} {\bibinfo {author} {\bibfnamefont {M.~H.~M.}\ \bibnamefont {Janssen}}, \bibinfo {author} {\bibfnamefont {M.}~\bibnamefont {Dantus}}, \bibinfo {author} {\bibfnamefont {H.}~\bibnamefont {guo}}, \ and\ \bibinfo {author} {\bibfnamefont {A.~H.}\ \bibnamefont {Zewail}},\ }\href {\doibase 10.1016/0009-2614(93)85635-2} {\bibfield  {journal} {\bibinfo  {journal} {Chemical Physics Letters}\ }\textbf {\bibinfo {volume} {214}},\ \bibinfo {pages} {281} (\bibinfo {year} {1993})}\BibitemShut {NoStop}%
\bibitem [{\citenamefont {Schwartz}\ and\ \citenamefont {Rossky}(1994)}]{Schwartz1994a}%
  \BibitemOpen
  \bibfield  {author} {\bibinfo {author} {\bibfnamefont {B.~J.}\ \bibnamefont {Schwartz}}\ and\ \bibinfo {author} {\bibfnamefont {P.~J.}\ \bibnamefont {Rossky}},\ }\href {\doibase 10.1021/j100068a001} {\bibfield  {journal} {\bibinfo  {journal} {The Journal of Physical Chemistry}\ }\textbf {\bibinfo {volume} {98}},\ \bibinfo {pages} {4489} (\bibinfo {year} {1994})}\BibitemShut {NoStop}%
\bibitem [{\citenamefont {Batista}\ and\ \citenamefont {Coker}(1997)}]{Batista1997}%
  \BibitemOpen
  \bibfield  {author} {\bibinfo {author} {\bibfnamefont {V.~S.}\ \bibnamefont {Batista}}\ and\ \bibinfo {author} {\bibfnamefont {D.~F.}\ \bibnamefont {Coker}},\ }\href {\doibase 10.1063/1.473717} {\bibfield  {journal} {\bibinfo  {journal} {The Journal of Chemical Physics}\ }\textbf {\bibinfo {volume} {106}},\ \bibinfo {pages} {6923} (\bibinfo {year} {1997})}\BibitemShut {NoStop}%
\bibitem [{\citenamefont {Vreven}\ \emph {et~al.}(1997)\citenamefont {Vreven}, \citenamefont {Bernardi}, \citenamefont {Garavelli}, \citenamefont {Olivucci}, \citenamefont {Robb},\ and\ \citenamefont {Schlegel}}]{Vreven1997}%
  \BibitemOpen
  \bibfield  {author} {\bibinfo {author} {\bibfnamefont {T.}~\bibnamefont {Vreven}}, \bibinfo {author} {\bibfnamefont {F.}~\bibnamefont {Bernardi}}, \bibinfo {author} {\bibfnamefont {M.}~\bibnamefont {Garavelli}}, \bibinfo {author} {\bibfnamefont {M.}~\bibnamefont {Olivucci}}, \bibinfo {author} {\bibfnamefont {M.}~\bibnamefont {Robb}}, \ and\ \bibinfo {author} {\bibfnamefont {H.}~\bibnamefont {Schlegel}},\ }\href {\doibase 10.1021/ja9725763} {\bibfield  {journal} {\bibinfo  {journal} {Journal of the American Chemical Society}\ }\textbf {\bibinfo {volume} {119}},\ \bibinfo {pages} {12687} (\bibinfo {year} {1997})}\BibitemShut {NoStop}%
\bibitem [{\citenamefont {Ben-Nun}\ and\ \citenamefont {Martinez}(1998)}]{BenNun1998}%
  \BibitemOpen
  \bibfield  {author} {\bibinfo {author} {\bibfnamefont {M.}~\bibnamefont {Ben-Nun}}\ and\ \bibinfo {author} {\bibfnamefont {T.}~\bibnamefont {Martinez}},\ }\href {\doibase 10.1016/S0009-2614(98)01115-4} {\bibfield  {journal} {\bibinfo  {journal} {Chemical Physics Letters}\ }\textbf {\bibinfo {volume} {298}},\ \bibinfo {pages} {57} (\bibinfo {year} {1998})}\BibitemShut {NoStop}%
\bibitem [{\citenamefont {Subotnik}\ \emph {et~al.}(2016)\citenamefont {Subotnik}, \citenamefont {Jain}, \citenamefont {Landry}, \citenamefont {Petit}, \citenamefont {Ouyang},\ and\ \citenamefont {Bellonzi}}]{Subotnik2016}%
  \BibitemOpen
  \bibfield  {author} {\bibinfo {author} {\bibfnamefont {J.~E.}\ \bibnamefont {Subotnik}}, \bibinfo {author} {\bibfnamefont {A.}~\bibnamefont {Jain}}, \bibinfo {author} {\bibfnamefont {B.}~\bibnamefont {Landry}}, \bibinfo {author} {\bibfnamefont {A.}~\bibnamefont {Petit}}, \bibinfo {author} {\bibfnamefont {W.}~\bibnamefont {Ouyang}}, \ and\ \bibinfo {author} {\bibfnamefont {N.}~\bibnamefont {Bellonzi}},\ }\href {\doibase 10.1146/annurev-physchem-040215-112245} {\bibfield  {journal} {\bibinfo  {journal} {Annual Review of Physical Chemistry}\ }\textbf {\bibinfo {volume} {67}},\ \bibinfo {pages} {387} (\bibinfo {year} {2016})},\ \bibinfo {note} {pMID: 27215818},\ \Eprint {http://arxiv.org/abs/https://doi.org/10.1146/annurev-physchem-040215-112245} {https://doi.org/10.1146/annurev-physchem-040215-112245} \BibitemShut {NoStop}%
\bibitem [{\citenamefont {Gossel}, \citenamefont {Agostini},\ and\ \citenamefont {Maitra}(2018)}]{Gossel2018}%
  \BibitemOpen
  \bibfield  {author} {\bibinfo {author} {\bibfnamefont {G.~H.}\ \bibnamefont {Gossel}}, \bibinfo {author} {\bibfnamefont {F.}~\bibnamefont {Agostini}}, \ and\ \bibinfo {author} {\bibfnamefont {N.~T.}\ \bibnamefont {Maitra}},\ }\href {\doibase 10.1021/acs.jctc.8b00449} {\bibfield  {journal} {\bibinfo  {journal} {Journal of Chemical Theory and Computation}\ }\textbf {\bibinfo {volume} {14}},\ \bibinfo {pages} {4513} (\bibinfo {year} {2018})}\BibitemShut {NoStop}%
\bibitem [{\citenamefont {Crespo-Otero}\ and\ \citenamefont {Barbatti}(2018)}]{barbatti_review}%
  \BibitemOpen
  \bibfield  {author} {\bibinfo {author} {\bibfnamefont {R.}~\bibnamefont {Crespo-Otero}}\ and\ \bibinfo {author} {\bibfnamefont {M.}~\bibnamefont {Barbatti}},\ }\href {\doibase 10.1021/acs.chemrev.7b00577} {\bibfield  {journal} {\bibinfo  {journal} {Chemical Reviews}\ }\textbf {\bibinfo {volume} {118}},\ \bibinfo {pages} {7026} (\bibinfo {year} {2018})},\ \bibinfo {note} {pMID: 29767966},\ \Eprint {http://arxiv.org/abs/https://doi.org/10.1021/acs.chemrev.7b00577} {https://doi.org/10.1021/acs.chemrev.7b00577} \BibitemShut {NoStop}%
\bibitem [{\citenamefont {Curchod}\ and\ \citenamefont {Martínez}(2018)}]{basiletodd_rev}%
  \BibitemOpen
  \bibfield  {author} {\bibinfo {author} {\bibfnamefont {B.~F.~E.}\ \bibnamefont {Curchod}}\ and\ \bibinfo {author} {\bibfnamefont {T.~J.}\ \bibnamefont {Martínez}},\ }\href {\doibase 10.1021/acs.chemrev.7b00423} {\bibfield  {journal} {\bibinfo  {journal} {Chemical Reviews}\ }\textbf {\bibinfo {volume} {118}},\ \bibinfo {pages} {3305} (\bibinfo {year} {2018})},\ \bibinfo {note} {pMID: 29465231},\ \Eprint {http://arxiv.org/abs/https://doi.org/10.1021/acs.chemrev.7b00423} {https://doi.org/10.1021/acs.chemrev.7b00423} \BibitemShut {NoStop}%
\bibitem [{\citenamefont {Gonzalez}\ and\ \citenamefont {Lindh}(2021)}]{gonzales_book}%
  \BibitemOpen
  \bibfield  {author} {\bibinfo {author} {\bibfnamefont {L.}~\bibnamefont {Gonzalez}}\ and\ \bibinfo {author} {\bibfnamefont {R.}~\bibnamefont {Lindh}},\ }\href@noop {} {\emph {\bibinfo {title} {Quantum chemistry and dynamics of excited states : methods and applications}}}\ (\bibinfo  {publisher} {John Wiley \& Sons, Ltd},\ \bibinfo {address} {Hoboken, NJ},\ \bibinfo {year} {2021})\BibitemShut {NoStop}%
\bibitem [{\citenamefont {Agostini}\ and\ \citenamefont {Gross}(2021)}]{Agostini2021}%
  \BibitemOpen
  \bibfield  {author} {\bibinfo {author} {\bibfnamefont {F.}~\bibnamefont {Agostini}}\ and\ \bibinfo {author} {\bibfnamefont {E.~K.~U.}\ \bibnamefont {Gross}},\ }\href {\doibase 10.1140/epjb/s10051-021-00171-2} {\bibfield  {journal} {\bibinfo  {journal} {European Physical Journal B}\ }\textbf {\bibinfo {volume} {94}} (\bibinfo {year} {2021}),\ 10.1140/epjb/s10051-021-00171-2}\BibitemShut {NoStop}%
\bibitem [{\citenamefont {Ibele}\ and\ \citenamefont {Curchod}(2020)}]{Ibele2020}%
  \BibitemOpen
  \bibfield  {author} {\bibinfo {author} {\bibfnamefont {L.~M.}\ \bibnamefont {Ibele}}\ and\ \bibinfo {author} {\bibfnamefont {B.~F.~E.}\ \bibnamefont {Curchod}},\ }\href {\doibase 10.1039/d0cp01353f} {\bibfield  {journal} {\bibinfo  {journal} {Physical Chemistry Chemical Physics}\ }\textbf {\bibinfo {volume} {22}},\ \bibinfo {pages} {15183} (\bibinfo {year} {2020})}\BibitemShut {NoStop}%
\bibitem [{\citenamefont {Zhang}\ and\ \citenamefont {Herbert}(2021)}]{Zhang2021}%
  \BibitemOpen
  \bibfield  {author} {\bibinfo {author} {\bibfnamefont {X.}~\bibnamefont {Zhang}}\ and\ \bibinfo {author} {\bibfnamefont {J.~M.}\ \bibnamefont {Herbert}},\ }\href {\doibase 10.1063/5.0062757} {\bibfield  {journal} {\bibinfo  {journal} {The Journal of Chemical Physics}\ }\textbf {\bibinfo {volume} {155}} (\bibinfo {year} {2021}),\ 10.1063/5.0062757}\BibitemShut {NoStop}%
\bibitem [{\citenamefont {Janos}\ and\ \citenamefont {Slavicek}(2023)}]{Janos2023}%
  \BibitemOpen
  \bibfield  {author} {\bibinfo {author} {\bibfnamefont {J.}~\bibnamefont {Janos}}\ and\ \bibinfo {author} {\bibfnamefont {P.}~\bibnamefont {Slavicek}},\ }\href {\doibase 10.1021/acs.jctc.3c00908} {\bibfield  {journal} {\bibinfo  {journal} {Journal of Chemical Theory and Computation}\ }\textbf {\bibinfo {volume} {19}},\ \bibinfo {pages} {8273} (\bibinfo {year} {2023})}\BibitemShut {NoStop}%
\bibitem [{\citenamefont {Akimov}(2016)}]{Akimov2016}%
  \BibitemOpen
  \bibfield  {author} {\bibinfo {author} {\bibfnamefont {A.~V.}\ \bibnamefont {Akimov}},\ }\href {\doibase 10.1002/jcc.24367} {\bibfield  {journal} {\bibinfo  {journal} {Journal of Computational Chemistry}\ }\textbf {\bibinfo {volume} {37}},\ \bibinfo {pages} {1626} (\bibinfo {year} {2016})}\BibitemShut {NoStop}%
\bibitem [{\citenamefont {Mai}, \citenamefont {Marquetand},\ and\ \citenamefont {Gonzalez}(2018)}]{mai2018}%
  \BibitemOpen
  \bibfield  {author} {\bibinfo {author} {\bibfnamefont {S.}~\bibnamefont {Mai}}, \bibinfo {author} {\bibfnamefont {P.}~\bibnamefont {Marquetand}}, \ and\ \bibinfo {author} {\bibfnamefont {L.}~\bibnamefont {Gonzalez}},\ }\href {\doibase 10.1002/wcms.1370} {\bibfield  {journal} {\bibinfo  {journal} {Wiley Interdisciplinary Reviews: Computational Molecular Science}\ }\textbf {\bibinfo {volume} {8}} (\bibinfo {year} {2018}),\ 10.1002/wcms.1370}\BibitemShut {NoStop}%
\bibitem [{\citenamefont {Fedorov}\ \emph {et~al.}(2020)\citenamefont {Fedorov}, \citenamefont {Seritan}, \citenamefont {Fales}, \citenamefont {Martinez},\ and\ \citenamefont {Levine}}]{fedorov2020}%
  \BibitemOpen
  \bibfield  {author} {\bibinfo {author} {\bibfnamefont {D.~A.}\ \bibnamefont {Fedorov}}, \bibinfo {author} {\bibfnamefont {S.}~\bibnamefont {Seritan}}, \bibinfo {author} {\bibfnamefont {B.~S.}\ \bibnamefont {Fales}}, \bibinfo {author} {\bibfnamefont {T.~J.}\ \bibnamefont {Martinez}}, \ and\ \bibinfo {author} {\bibfnamefont {B.~G.}\ \bibnamefont {Levine}},\ }\href {\doibase 10.1021/acs.jctc.0c00575} {\bibfield  {journal} {\bibinfo  {journal} {Journal of Chemical Theory and Computation}\ }\textbf {\bibinfo {volume} {16}},\ \bibinfo {pages} {5485} (\bibinfo {year} {2020})}\BibitemShut {NoStop}%
\bibitem [{\citenamefont {Malone}\ \emph {et~al.}(2020)\citenamefont {Malone}, \citenamefont {Nebgen}, \citenamefont {White}, \citenamefont {Zhang}, \citenamefont {Song}, \citenamefont {Bjorgaard}, \citenamefont {Sifain}, \citenamefont {Rodriguez-Hernandez}, \citenamefont {Freixas}, \citenamefont {Fernandez-Alberti}, \citenamefont {Roitberg}, \citenamefont {Nelson},\ and\ \citenamefont {Tretiak}}]{malone2020}%
  \BibitemOpen
  \bibfield  {author} {\bibinfo {author} {\bibfnamefont {W.}~\bibnamefont {Malone}}, \bibinfo {author} {\bibfnamefont {B.}~\bibnamefont {Nebgen}}, \bibinfo {author} {\bibfnamefont {A.}~\bibnamefont {White}}, \bibinfo {author} {\bibfnamefont {Y.}~\bibnamefont {Zhang}}, \bibinfo {author} {\bibfnamefont {H.}~\bibnamefont {Song}}, \bibinfo {author} {\bibfnamefont {J.~A.}\ \bibnamefont {Bjorgaard}}, \bibinfo {author} {\bibfnamefont {A.~E.}\ \bibnamefont {Sifain}}, \bibinfo {author} {\bibfnamefont {B.}~\bibnamefont {Rodriguez-Hernandez}}, \bibinfo {author} {\bibfnamefont {V.~M.}\ \bibnamefont {Freixas}}, \bibinfo {author} {\bibfnamefont {S.}~\bibnamefont {Fernandez-Alberti}}, \bibinfo {author} {\bibfnamefont {A.~E.}\ \bibnamefont {Roitberg}}, \bibinfo {author} {\bibfnamefont {T.~R.}\ \bibnamefont {Nelson}}, \ and\ \bibinfo {author} {\bibfnamefont {S.}~\bibnamefont {Tretiak}},\ }\href {\doibase 10.1021/acs.jctc.0c00248} {\bibfield  {journal} {\bibinfo  {journal} {Journal of Chemical Theory and Computation}\ }\textbf
  {\bibinfo {volume} {16}},\ \bibinfo {pages} {5771} (\bibinfo {year} {2020})}\BibitemShut {NoStop}%
\bibitem [{\citenamefont {Barbatti}\ \emph {et~al.}(2022)\citenamefont {Barbatti}, \citenamefont {Bondanza}, \citenamefont {Crespo-Otero}, \citenamefont {Demoulin}, \citenamefont {Dral}, \citenamefont {Granucci}, \citenamefont {Kossoski}, \citenamefont {Lischka}, \citenamefont {Mennucci}, \citenamefont {Mukherjee}, \citenamefont {Pederzoli}, \citenamefont {Persico}, \citenamefont {Pinheiro}, \citenamefont {Pittner}, \citenamefont {Plasser}, \citenamefont {Gil},\ and\ \citenamefont {Stojanovic}}]{Barbatti2022}%
  \BibitemOpen
  \bibfield  {author} {\bibinfo {author} {\bibfnamefont {M.}~\bibnamefont {Barbatti}}, \bibinfo {author} {\bibfnamefont {M.}~\bibnamefont {Bondanza}}, \bibinfo {author} {\bibfnamefont {R.}~\bibnamefont {Crespo-Otero}}, \bibinfo {author} {\bibfnamefont {B.}~\bibnamefont {Demoulin}}, \bibinfo {author} {\bibfnamefont {P.~O.}\ \bibnamefont {Dral}}, \bibinfo {author} {\bibfnamefont {G.}~\bibnamefont {Granucci}}, \bibinfo {author} {\bibfnamefont {F.}~\bibnamefont {Kossoski}}, \bibinfo {author} {\bibfnamefont {H.}~\bibnamefont {Lischka}}, \bibinfo {author} {\bibfnamefont {B.}~\bibnamefont {Mennucci}}, \bibinfo {author} {\bibfnamefont {S.}~\bibnamefont {Mukherjee}}, \bibinfo {author} {\bibfnamefont {M.}~\bibnamefont {Pederzoli}}, \bibinfo {author} {\bibfnamefont {M.}~\bibnamefont {Persico}}, \bibinfo {author} {\bibfnamefont {M.}~\bibnamefont {Pinheiro}}, \bibinfo {author} {\bibfnamefont {J.}~\bibnamefont {Pittner}}, \bibinfo {author} {\bibfnamefont {F.}~\bibnamefont {Plasser}}, \bibinfo {author} {\bibfnamefont
  {E.~S.}\ \bibnamefont {Gil}}, \ and\ \bibinfo {author} {\bibfnamefont {L.}~\bibnamefont {Stojanovic}},\ }\href {\doibase 10.1021/acs.jctc.2c00804} {\bibfield  {journal} {\bibinfo  {journal} {Journal of Chemical Theory and Computation}\ }\textbf {\bibinfo {volume} {18}},\ \bibinfo {pages} {6851} (\bibinfo {year} {2022})}\BibitemShut {NoStop}%
\bibitem [{\citenamefont {Barbatti}\ \emph {et~al.}(2010)\citenamefont {Barbatti}, \citenamefont {Aquino}, \citenamefont {Szymczak}, \citenamefont {Nachtigallova}, \citenamefont {Hobza},\ and\ \citenamefont {Lischka}}]{Barbatti2010}%
  \BibitemOpen
  \bibfield  {author} {\bibinfo {author} {\bibfnamefont {M.}~\bibnamefont {Barbatti}}, \bibinfo {author} {\bibfnamefont {A.~J.~A.}\ \bibnamefont {Aquino}}, \bibinfo {author} {\bibfnamefont {J.~J.}\ \bibnamefont {Szymczak}}, \bibinfo {author} {\bibfnamefont {D.}~\bibnamefont {Nachtigallova}}, \bibinfo {author} {\bibfnamefont {P.}~\bibnamefont {Hobza}}, \ and\ \bibinfo {author} {\bibfnamefont {H.}~\bibnamefont {Lischka}},\ }\href {\doibase 10.1073/pnas.1014982107} {\bibfield  {journal} {\bibinfo  {journal} {Proceedings of the National Academy of Sciences}\ }\textbf {\bibinfo {volume} {107}},\ \bibinfo {pages} {21453} (\bibinfo {year} {2010})}\BibitemShut {NoStop}%
\bibitem [{\citenamefont {Richter}\ \emph {et~al.}(2012)\citenamefont {Richter}, \citenamefont {Marquetand}, \citenamefont {Gonzalez-Vazquez}, \citenamefont {Sola},\ and\ \citenamefont {Gonzalez}}]{Richter2012}%
  \BibitemOpen
  \bibfield  {author} {\bibinfo {author} {\bibfnamefont {M.}~\bibnamefont {Richter}}, \bibinfo {author} {\bibfnamefont {P.}~\bibnamefont {Marquetand}}, \bibinfo {author} {\bibfnamefont {J.}~\bibnamefont {Gonzalez-Vazquez}}, \bibinfo {author} {\bibfnamefont {I.}~\bibnamefont {Sola}}, \ and\ \bibinfo {author} {\bibfnamefont {L.}~\bibnamefont {Gonzalez}},\ }\href {\doibase 10.1021/jz301312h} {\bibfield  {journal} {\bibinfo  {journal} {The Journal of Physical Chemistry Letters}\ }\textbf {\bibinfo {volume} {3}},\ \bibinfo {pages} {3090} (\bibinfo {year} {2012})}\BibitemShut {NoStop}%
\bibitem [{\citenamefont {Penfold}\ \emph {et~al.}(2012)\citenamefont {Penfold}, \citenamefont {Spesyvtsev}, \citenamefont {Kirkby}, \citenamefont {Minns}, \citenamefont {Parker}, \citenamefont {Fielding},\ and\ \citenamefont {Worth}}]{Penfold2012}%
  \BibitemOpen
  \bibfield  {author} {\bibinfo {author} {\bibfnamefont {T.~J.}\ \bibnamefont {Penfold}}, \bibinfo {author} {\bibfnamefont {R.}~\bibnamefont {Spesyvtsev}}, \bibinfo {author} {\bibfnamefont {O.~M.}\ \bibnamefont {Kirkby}}, \bibinfo {author} {\bibfnamefont {R.~S.}\ \bibnamefont {Minns}}, \bibinfo {author} {\bibfnamefont {D.~S.~N.}\ \bibnamefont {Parker}}, \bibinfo {author} {\bibfnamefont {H.~H.}\ \bibnamefont {Fielding}}, \ and\ \bibinfo {author} {\bibfnamefont {G.~A.}\ \bibnamefont {Worth}},\ }\href {\doibase 10.1063/1.4767054} {\bibfield  {journal} {\bibinfo  {journal} {The Journal of Chemical Physics}\ }\textbf {\bibinfo {volume} {137}} (\bibinfo {year} {2012}),\ 10.1063/1.4767054}\BibitemShut {NoStop}%
\bibitem [{\citenamefont {Nelson}\ \emph {et~al.}(2014)\citenamefont {Nelson}, \citenamefont {Fernandez-Alberti}, \citenamefont {Roitberg},\ and\ \citenamefont {Tretiak}}]{nelson2014}%
  \BibitemOpen
  \bibfield  {author} {\bibinfo {author} {\bibfnamefont {T.}~\bibnamefont {Nelson}}, \bibinfo {author} {\bibfnamefont {S.}~\bibnamefont {Fernandez-Alberti}}, \bibinfo {author} {\bibfnamefont {A.~E.}\ \bibnamefont {Roitberg}}, \ and\ \bibinfo {author} {\bibfnamefont {S.}~\bibnamefont {Tretiak}},\ }\href {\doibase 10.1021/ar400263p} {\bibfield  {journal} {\bibinfo  {journal} {Accounts of Chemical Research}\ }\textbf {\bibinfo {volume} {47}},\ \bibinfo {pages} {1155} (\bibinfo {year} {2014})}\BibitemShut {NoStop}%
\bibitem [{\citenamefont {Tavernelli}(2015)}]{tavernelli2015}%
  \BibitemOpen
  \bibfield  {author} {\bibinfo {author} {\bibfnamefont {I.}~\bibnamefont {Tavernelli}},\ }\href {\doibase 10.1021/ar500357y} {\bibfield  {journal} {\bibinfo  {journal} {Accounts of Chemical Research}\ }\textbf {\bibinfo {volume} {48}},\ \bibinfo {pages} {792} (\bibinfo {year} {2015})}\BibitemShut {NoStop}%
\bibitem [{\citenamefont {Wang}, \citenamefont {Long},\ and\ \citenamefont {Prezhdo}(2015)}]{wang2015}%
  \BibitemOpen
  \bibfield  {author} {\bibinfo {author} {\bibfnamefont {L.}~\bibnamefont {Wang}}, \bibinfo {author} {\bibfnamefont {R.}~\bibnamefont {Long}}, \ and\ \bibinfo {author} {\bibfnamefont {O.~V.}\ \bibnamefont {Prezhdo}},\ }\href {\doibase 10.1146/annurev-physchem-040214-121359} {\bibfield  {journal} {\bibinfo  {journal} {Annual Review of Physical Chemistry}\ }\textbf {\bibinfo {volume} {66}},\ \bibinfo {pages} {549} (\bibinfo {year} {2015})},\ \bibinfo {note} {pMID: 25622188},\ \Eprint {http://arxiv.org/abs/https://doi.org/10.1146/annurev-physchem-040214-121359} {https://doi.org/10.1146/annurev-physchem-040214-121359} \BibitemShut {NoStop}%
\bibitem [{\citenamefont {Fielding}\ and\ \citenamefont {Worth}(2018)}]{Fielding2018}%
  \BibitemOpen
  \bibfield  {author} {\bibinfo {author} {\bibfnamefont {H.~H.}\ \bibnamefont {Fielding}}\ and\ \bibinfo {author} {\bibfnamefont {G.~A.}\ \bibnamefont {Worth}},\ }\href {\doibase 10.1039/c7cs00627f} {\bibfield  {journal} {\bibinfo  {journal} {Chemical Society Reviews}\ }\textbf {\bibinfo {volume} {47}},\ \bibinfo {pages} {309} (\bibinfo {year} {2018})}\BibitemShut {NoStop}%
\bibitem [{\citenamefont {Schuurman}\ and\ \citenamefont {Stolow}(2018)}]{schuurman2018}%
  \BibitemOpen
  \bibfield  {author} {\bibinfo {author} {\bibfnamefont {M.~S.}\ \bibnamefont {Schuurman}}\ and\ \bibinfo {author} {\bibfnamefont {A.}~\bibnamefont {Stolow}},\ }\href {\doibase 10.1146/annurev-physchem-052516-050721} {\bibfield  {journal} {\bibinfo  {journal} {Annual Review of Physical Chemistry}\ }\textbf {\bibinfo {volume} {69}},\ \bibinfo {pages} {427} (\bibinfo {year} {2018})},\ \bibinfo {note} {pMID: 29490199},\ \Eprint {http://arxiv.org/abs/https://doi.org/10.1146/annurev-physchem-052516-050721} {https://doi.org/10.1146/annurev-physchem-052516-050721} \BibitemShut {NoStop}%
\bibitem [{\citenamefont {Popp}\ \emph {et~al.}(2021)\citenamefont {Popp}, \citenamefont {Brey}, \citenamefont {Binder},\ and\ \citenamefont {Burghardt}}]{popp2021}%
  \BibitemOpen
  \bibfield  {author} {\bibinfo {author} {\bibfnamefont {W.}~\bibnamefont {Popp}}, \bibinfo {author} {\bibfnamefont {D.}~\bibnamefont {Brey}}, \bibinfo {author} {\bibfnamefont {R.}~\bibnamefont {Binder}}, \ and\ \bibinfo {author} {\bibfnamefont {I.}~\bibnamefont {Burghardt}},\ }\href {\doibase 10.1146/annurev-physchem-090419-040306} {\bibfield  {journal} {\bibinfo  {journal} {Annual Review of Physical Chemistry}\ }\textbf {\bibinfo {volume} {72}},\ \bibinfo {pages} {591} (\bibinfo {year} {2021})},\ \bibinfo {note} {pMID: 33636997}\BibitemShut {NoStop}%
\bibitem [{\citenamefont {Talotta}, \citenamefont {Lauvergnat},\ and\ \citenamefont {Agostini}(2022)}]{Talotta2022}%
  \BibitemOpen
  \bibfield  {author} {\bibinfo {author} {\bibfnamefont {F.}~\bibnamefont {Talotta}}, \bibinfo {author} {\bibfnamefont {D.}~\bibnamefont {Lauvergnat}}, \ and\ \bibinfo {author} {\bibfnamefont {F.}~\bibnamefont {Agostini}},\ }\href {\doibase 10.1063/5.0089415} {\bibfield  {journal} {\bibinfo  {journal} {The Journal of Chemical Physics}\ }\textbf {\bibinfo {volume} {156}} (\bibinfo {year} {2022}),\ 10.1063/5.0089415}\BibitemShut {NoStop}%
\bibitem [{\citenamefont {Dergachev}\ \emph {et~al.}(2023)\citenamefont {Dergachev}, \citenamefont {Dergachev}, \citenamefont {Rooein}, \citenamefont {Mirzanejad},\ and\ \citenamefont {Varganov}}]{dergachev2023}%
  \BibitemOpen
  \bibfield  {author} {\bibinfo {author} {\bibfnamefont {I.~D.}\ \bibnamefont {Dergachev}}, \bibinfo {author} {\bibfnamefont {V.~D.}\ \bibnamefont {Dergachev}}, \bibinfo {author} {\bibfnamefont {M.}~\bibnamefont {Rooein}}, \bibinfo {author} {\bibfnamefont {A.}~\bibnamefont {Mirzanejad}}, \ and\ \bibinfo {author} {\bibfnamefont {S.~A.}\ \bibnamefont {Varganov}},\ }\href {\doibase 10.1021/acs.accounts.2c00843} {\bibfield  {journal} {\bibinfo  {journal} {Accounts of Chemical Research}\ }\textbf {\bibinfo {volume} {56}},\ \bibinfo {pages} {856} (\bibinfo {year} {2023})}\BibitemShut {NoStop}%
\bibitem [{\citenamefont {Hudock}\ \emph {et~al.}(2007)\citenamefont {Hudock}, \citenamefont {Levine}, \citenamefont {Thompson}, \citenamefont {Satzger}, \citenamefont {Townsend}, \citenamefont {Gador}, \citenamefont {Ullrich}, \citenamefont {Stolow},\ and\ \citenamefont {Martínez}}]{Hudock2007}%
  \BibitemOpen
  \bibfield  {author} {\bibinfo {author} {\bibfnamefont {H.~R.}\ \bibnamefont {Hudock}}, \bibinfo {author} {\bibfnamefont {B.~G.}\ \bibnamefont {Levine}}, \bibinfo {author} {\bibfnamefont {A.~L.}\ \bibnamefont {Thompson}}, \bibinfo {author} {\bibfnamefont {H.}~\bibnamefont {Satzger}}, \bibinfo {author} {\bibfnamefont {D.}~\bibnamefont {Townsend}}, \bibinfo {author} {\bibfnamefont {N.}~\bibnamefont {Gador}}, \bibinfo {author} {\bibfnamefont {S.}~\bibnamefont {Ullrich}}, \bibinfo {author} {\bibfnamefont {A.}~\bibnamefont {Stolow}}, \ and\ \bibinfo {author} {\bibfnamefont {T.~J.}\ \bibnamefont {Martínez}},\ }\href {\doibase 10.1021/jp0723665} {\bibfield  {journal} {\bibinfo  {journal} {The Journal of Physical Chemistry A}\ }\textbf {\bibinfo {volume} {111}},\ \bibinfo {pages} {8500} (\bibinfo {year} {2007})},\ \bibinfo {note} {pMID: 17685594},\ \Eprint {http://arxiv.org/abs/https://doi.org/10.1021/jp0723665} {https://doi.org/10.1021/jp0723665} \BibitemShut {NoStop}%
\bibitem [{\citenamefont {Mitric}\ \emph {et~al.}(2011)\citenamefont {Mitric}, \citenamefont {Petersen}, \citenamefont {Wohlgemuth}, \citenamefont {Werner}, \citenamefont {Bonacic-Koutecky}, \citenamefont {Woeste},\ and\ \citenamefont {Jortner}}]{Mitric2011}%
  \BibitemOpen
  \bibfield  {author} {\bibinfo {author} {\bibfnamefont {R.}~\bibnamefont {Mitric}}, \bibinfo {author} {\bibfnamefont {J.}~\bibnamefont {Petersen}}, \bibinfo {author} {\bibfnamefont {M.}~\bibnamefont {Wohlgemuth}}, \bibinfo {author} {\bibfnamefont {U.}~\bibnamefont {Werner}}, \bibinfo {author} {\bibfnamefont {V.}~\bibnamefont {Bonacic-Koutecky}}, \bibinfo {author} {\bibfnamefont {L.}~\bibnamefont {Woeste}}, \ and\ \bibinfo {author} {\bibfnamefont {J.}~\bibnamefont {Jortner}},\ }\href {\doibase 10.1021/jp106355n} {\bibfield  {journal} {\bibinfo  {journal} {The Journal of Physical Chemistry A}\ }\textbf {\bibinfo {volume} {115}},\ \bibinfo {pages} {3755} (\bibinfo {year} {2011})}\BibitemShut {NoStop}%
\bibitem [{\citenamefont {De~Giovannini}\ \emph {et~al.}(2013)\citenamefont {De~Giovannini}, \citenamefont {Brunetto}, \citenamefont {Castro}, \citenamefont {Walkenhorst},\ and\ \citenamefont {Rubio}}]{Rubio2013}%
  \BibitemOpen
  \bibfield  {author} {\bibinfo {author} {\bibfnamefont {U.}~\bibnamefont {De~Giovannini}}, \bibinfo {author} {\bibfnamefont {G.}~\bibnamefont {Brunetto}}, \bibinfo {author} {\bibfnamefont {A.}~\bibnamefont {Castro}}, \bibinfo {author} {\bibfnamefont {J.}~\bibnamefont {Walkenhorst}}, \ and\ \bibinfo {author} {\bibfnamefont {A.}~\bibnamefont {Rubio}},\ }\href {\doibase https://doi.org/10.1002/cphc.201201007} {\bibfield  {journal} {\bibinfo  {journal} {ChemPhysChem}\ }\textbf {\bibinfo {volume} {14}},\ \bibinfo {pages} {1363} (\bibinfo {year} {2013})}\BibitemShut {NoStop}%
\bibitem [{\citenamefont {Petit}\ and\ \citenamefont {Subotnik}(2014)}]{Subotnik2014}%
  \BibitemOpen
  \bibfield  {author} {\bibinfo {author} {\bibfnamefont {A.~S.}\ \bibnamefont {Petit}}\ and\ \bibinfo {author} {\bibfnamefont {J.~E.}\ \bibnamefont {Subotnik}},\ }\href {\doibase 10.1063/1.4897258} {\bibfield  {journal} {\bibinfo  {journal} {The Journal of Chemical Physics}\ }\textbf {\bibinfo {volume} {141}},\ \bibinfo {pages} {154108} (\bibinfo {year} {2014})}\BibitemShut {NoStop}%
\bibitem [{\citenamefont {Fischer}, \citenamefont {Cramer},\ and\ \citenamefont {Govind}(2015)}]{Govind2015}%
  \BibitemOpen
  \bibfield  {author} {\bibinfo {author} {\bibfnamefont {S.~A.}\ \bibnamefont {Fischer}}, \bibinfo {author} {\bibfnamefont {C.~J.}\ \bibnamefont {Cramer}}, \ and\ \bibinfo {author} {\bibfnamefont {N.}~\bibnamefont {Govind}},\ }\href {\doibase 10.1021/acs.jctc.5b00473} {\bibfield  {journal} {\bibinfo  {journal} {Journal of Chemical Theory and Computation}\ }\textbf {\bibinfo {volume} {11}},\ \bibinfo {pages} {4294} (\bibinfo {year} {2015})}\BibitemShut {NoStop}%
\bibitem [{\citenamefont {Nguyen}\ \emph {et~al.}(2016)\citenamefont {Nguyen}, \citenamefont {Koh}, \citenamefont {Lefelhocz},\ and\ \citenamefont {Parkhill}}]{Parkhill2016}%
  \BibitemOpen
  \bibfield  {author} {\bibinfo {author} {\bibfnamefont {T.~S.}\ \bibnamefont {Nguyen}}, \bibinfo {author} {\bibfnamefont {J.~H.}\ \bibnamefont {Koh}}, \bibinfo {author} {\bibfnamefont {S.}~\bibnamefont {Lefelhocz}}, \ and\ \bibinfo {author} {\bibfnamefont {J.}~\bibnamefont {Parkhill}},\ }\href {\doibase 10.1021/acs.jpclett.6b00421} {\bibfield  {journal} {\bibinfo  {journal} {The Journal of Physical Chemistry Letters}\ }\textbf {\bibinfo {volume} {7}},\ \bibinfo {pages} {1590} (\bibinfo {year} {2016})}\BibitemShut {NoStop}%
\bibitem [{\citenamefont {Dsouza}\ \emph {et~al.}(2018)\citenamefont {Dsouza}, \citenamefont {Cheng}, \citenamefont {Li}, \citenamefont {Miller},\ and\ \citenamefont {Kochman}}]{Dsouza2018}%
  \BibitemOpen
  \bibfield  {author} {\bibinfo {author} {\bibfnamefont {R.}~\bibnamefont {Dsouza}}, \bibinfo {author} {\bibfnamefont {X.}~\bibnamefont {Cheng}}, \bibinfo {author} {\bibfnamefont {Z.}~\bibnamefont {Li}}, \bibinfo {author} {\bibfnamefont {R.~J.~D.}\ \bibnamefont {Miller}}, \ and\ \bibinfo {author} {\bibfnamefont {M.~A.}\ \bibnamefont {Kochman}},\ }\href {\doibase 10.1021/acs.jpca.8b10241} {\bibfield  {journal} {\bibinfo  {journal} {The Journal of Physical Chemistry A}\ }\textbf {\bibinfo {volume} {122}},\ \bibinfo {pages} {9688} (\bibinfo {year} {2018})}\BibitemShut {NoStop}%
\bibitem [{\citenamefont {Bonafé}\ \emph {et~al.}(2018)\citenamefont {Bonafé}, \citenamefont {Hernández}, \citenamefont {Aradi}, \citenamefont {Frauenheim},\ and\ \citenamefont {Sánchez}}]{Sanchez2018}%
  \BibitemOpen
  \bibfield  {author} {\bibinfo {author} {\bibfnamefont {F.~P.}\ \bibnamefont {Bonafé}}, \bibinfo {author} {\bibfnamefont {F.~J.}\ \bibnamefont {Hernández}}, \bibinfo {author} {\bibfnamefont {B.}~\bibnamefont {Aradi}}, \bibinfo {author} {\bibfnamefont {T.}~\bibnamefont {Frauenheim}}, \ and\ \bibinfo {author} {\bibfnamefont {C.~G.}\ \bibnamefont {Sánchez}},\ }\href {\doibase 10.1021/acs.jpclett.8b01659} {\bibfield  {journal} {\bibinfo  {journal} {The Journal of Physical Chemistry Letters}\ }\textbf {\bibinfo {volume} {9}},\ \bibinfo {pages} {4355} (\bibinfo {year} {2018})}\BibitemShut {NoStop}%
\bibitem [{\citenamefont {Gelin}\ \emph {et~al.}(2021)\citenamefont {Gelin}, \citenamefont {Huang}, \citenamefont {Xie}, \citenamefont {Chen}, \citenamefont {Došlić},\ and\ \citenamefont {Domcke}}]{Domcke2021}%
  \BibitemOpen
  \bibfield  {author} {\bibinfo {author} {\bibfnamefont {M.~F.}\ \bibnamefont {Gelin}}, \bibinfo {author} {\bibfnamefont {X.}~\bibnamefont {Huang}}, \bibinfo {author} {\bibfnamefont {W.}~\bibnamefont {Xie}}, \bibinfo {author} {\bibfnamefont {L.}~\bibnamefont {Chen}}, \bibinfo {author} {\bibfnamefont {N.}~\bibnamefont {Došlić}}, \ and\ \bibinfo {author} {\bibfnamefont {W.}~\bibnamefont {Domcke}},\ }\href {\doibase 10.1021/acs.jctc.1c00109} {\bibfield  {journal} {\bibinfo  {journal} {Journal of Chemical Theory and Computation}\ }\textbf {\bibinfo {volume} {17}},\ \bibinfo {pages} {2394} (\bibinfo {year} {2021})}\BibitemShut {NoStop}%
\bibitem [{\citenamefont {Kochman}, \citenamefont {Durbeej},\ and\ \citenamefont {Kubas}(2021)}]{Kubas2021}%
  \BibitemOpen
  \bibfield  {author} {\bibinfo {author} {\bibfnamefont {M.~A.}\ \bibnamefont {Kochman}}, \bibinfo {author} {\bibfnamefont {B.}~\bibnamefont {Durbeej}}, \ and\ \bibinfo {author} {\bibfnamefont {A.}~\bibnamefont {Kubas}},\ }\href {\doibase 10.1021/acs.jpca.1c06166} {\bibfield  {journal} {\bibinfo  {journal} {The Journal of Physical Chemistry A}\ }\textbf {\bibinfo {volume} {125}},\ \bibinfo {pages} {8635} (\bibinfo {year} {2021})}\BibitemShut {NoStop}%
\bibitem [{\citenamefont {Borrego-Varillas}\ \emph {et~al.}(2021)\citenamefont {Borrego-Varillas}, \citenamefont {Nenov}, \citenamefont {Kabaciński}, \citenamefont {Conti}, \citenamefont {Ganzer}, \citenamefont {Oriana}, \citenamefont {Jaiswal}, \citenamefont {Delfino}, \citenamefont {Weingart}, \citenamefont {Manzoni}, \citenamefont {Rivalta}, \citenamefont {Garavelli},\ and\ \citenamefont {Cerullo}}]{Cerullo2021}%
  \BibitemOpen
  \bibfield  {author} {\bibinfo {author} {\bibfnamefont {R.}~\bibnamefont {Borrego-Varillas}}, \bibinfo {author} {\bibfnamefont {A.}~\bibnamefont {Nenov}}, \bibinfo {author} {\bibfnamefont {P.}~\bibnamefont {Kabaciński}}, \bibinfo {author} {\bibfnamefont {I.}~\bibnamefont {Conti}}, \bibinfo {author} {\bibfnamefont {L.}~\bibnamefont {Ganzer}}, \bibinfo {author} {\bibfnamefont {A.}~\bibnamefont {Oriana}}, \bibinfo {author} {\bibfnamefont {V.~K.}\ \bibnamefont {Jaiswal}}, \bibinfo {author} {\bibfnamefont {I.}~\bibnamefont {Delfino}}, \bibinfo {author} {\bibfnamefont {O.}~\bibnamefont {Weingart}}, \bibinfo {author} {\bibfnamefont {C.}~\bibnamefont {Manzoni}}, \bibinfo {author} {\bibfnamefont {I.}~\bibnamefont {Rivalta}}, \bibinfo {author} {\bibfnamefont {M.}~\bibnamefont {Garavelli}}, \ and\ \bibinfo {author} {\bibfnamefont {G.}~\bibnamefont {Cerullo}},\ }\href {\doibase 10.1038/s41467-021-27535-7} {\bibfield  {journal} {\bibinfo  {journal} {Nature Communications}\ }\textbf {\bibinfo {volume} {12}},\ \bibinfo
  {pages} {7285} (\bibinfo {year} {2021})}\BibitemShut {NoStop}%
\bibitem [{\citenamefont {Xu}\ \emph {et~al.}(2022)\citenamefont {Xu}, \citenamefont {Lin}, \citenamefont {Hu}, \citenamefont {Gu}, \citenamefont {Gelin},\ and\ \citenamefont {Lan}}]{Xu2022}%
  \BibitemOpen
  \bibfield  {author} {\bibinfo {author} {\bibfnamefont {C.}~\bibnamefont {Xu}}, \bibinfo {author} {\bibfnamefont {K.}~\bibnamefont {Lin}}, \bibinfo {author} {\bibfnamefont {D.}~\bibnamefont {Hu}}, \bibinfo {author} {\bibfnamefont {F.~L.}\ \bibnamefont {Gu}}, \bibinfo {author} {\bibfnamefont {M.~F.}\ \bibnamefont {Gelin}}, \ and\ \bibinfo {author} {\bibfnamefont {Z.}~\bibnamefont {Lan}},\ }\href {\doibase 10.1021/acs.jpclett.1c03373} {\bibfield  {journal} {\bibinfo  {journal} {The Journal of Physical Chemistry Letters}\ }\textbf {\bibinfo {volume} {13}},\ \bibinfo {pages} {661} (\bibinfo {year} {2022})},\ \Eprint {http://arxiv.org/abs/https://doi.org/10.1021/acs.jpclett.1c03373} {https://doi.org/10.1021/acs.jpclett.1c03373} \BibitemShut {NoStop}%
\bibitem [{\citenamefont {Chakraborty}\ \emph {et~al.}(2022)\citenamefont {Chakraborty}, \citenamefont {Liu}, \citenamefont {McClung}, \citenamefont {Weinacht},\ and\ \citenamefont {Matsika}}]{Chakraborty2022}%
  \BibitemOpen
  \bibfield  {author} {\bibinfo {author} {\bibfnamefont {P.}~\bibnamefont {Chakraborty}}, \bibinfo {author} {\bibfnamefont {Y.}~\bibnamefont {Liu}}, \bibinfo {author} {\bibfnamefont {S.}~\bibnamefont {McClung}}, \bibinfo {author} {\bibfnamefont {T.}~\bibnamefont {Weinacht}}, \ and\ \bibinfo {author} {\bibfnamefont {S.}~\bibnamefont {Matsika}},\ }\href {\doibase 10.1021/acs.jpca.2c04671} {\bibfield  {journal} {\bibinfo  {journal} {The Journal of Physical Chemistry A}\ } (\bibinfo {year} {2022}),\ 10.1021/acs.jpca.2c04671}\BibitemShut {NoStop}%
\bibitem [{\citenamefont {Silfies}\ \emph {et~al.}(2023)\citenamefont {Silfies}, \citenamefont {Mehmood}, \citenamefont {Kowzan}, \citenamefont {Hohenstein}, \citenamefont {Levine},\ and\ \citenamefont {Allison}}]{Silfies2023}%
  \BibitemOpen
  \bibfield  {author} {\bibinfo {author} {\bibfnamefont {M.~C.}\ \bibnamefont {Silfies}}, \bibinfo {author} {\bibfnamefont {A.}~\bibnamefont {Mehmood}}, \bibinfo {author} {\bibfnamefont {G.}~\bibnamefont {Kowzan}}, \bibinfo {author} {\bibfnamefont {E.~G.}\ \bibnamefont {Hohenstein}}, \bibinfo {author} {\bibfnamefont {B.~G.}\ \bibnamefont {Levine}}, \ and\ \bibinfo {author} {\bibfnamefont {T.~K.}\ \bibnamefont {Allison}},\ }\href {\doibase 10.1063/5.0161238} {\bibfield  {journal} {\bibinfo  {journal} {The Journal of Chemical Physics}\ }\textbf {\bibinfo {volume} {159}} (\bibinfo {year} {2023}),\ 10.1063/5.0161238}\BibitemShut {NoStop}%
\bibitem [{\citenamefont {Young}\ \emph {et~al.}(2018)\citenamefont {Young}, \citenamefont {Ueda}, \citenamefont {G{\"u}hr}, \citenamefont {Bucksbaum}, \citenamefont {Simon}, \citenamefont {Mukamel}, \citenamefont {Rohringer}, \citenamefont {Prince}, \citenamefont {Masciovecchio}, \citenamefont {Meyer} \emph {et~al.}}]{young2018roadmap}%
  \BibitemOpen
  \bibfield  {author} {\bibinfo {author} {\bibfnamefont {L.}~\bibnamefont {Young}}, \bibinfo {author} {\bibfnamefont {K.}~\bibnamefont {Ueda}}, \bibinfo {author} {\bibfnamefont {M.}~\bibnamefont {G{\"u}hr}}, \bibinfo {author} {\bibfnamefont {P.~H.}\ \bibnamefont {Bucksbaum}}, \bibinfo {author} {\bibfnamefont {M.}~\bibnamefont {Simon}}, \bibinfo {author} {\bibfnamefont {S.}~\bibnamefont {Mukamel}}, \bibinfo {author} {\bibfnamefont {N.}~\bibnamefont {Rohringer}}, \bibinfo {author} {\bibfnamefont {K.~C.}\ \bibnamefont {Prince}}, \bibinfo {author} {\bibfnamefont {C.}~\bibnamefont {Masciovecchio}}, \bibinfo {author} {\bibfnamefont {M.}~\bibnamefont {Meyer}},  \emph {et~al.},\ }\href {\doibase 10.1088/1361-6455/aa9735} {\bibfield  {journal} {\bibinfo  {journal} {Journal of Physics B: Atomic, Molecular and Optical Physics}\ }\textbf {\bibinfo {volume} {51}},\ \bibinfo {pages} {032003} (\bibinfo {year} {2018})}\BibitemShut {NoStop}%
\bibitem [{\citenamefont {Nisoli}\ \emph {et~al.}(2017)\citenamefont {Nisoli}, \citenamefont {Decleva}, \citenamefont {Calegari}, \citenamefont {Palacios},\ and\ \citenamefont {Mart{\'\i}n}}]{nisoli2017attosecond}%
  \BibitemOpen
  \bibfield  {author} {\bibinfo {author} {\bibfnamefont {M.}~\bibnamefont {Nisoli}}, \bibinfo {author} {\bibfnamefont {P.}~\bibnamefont {Decleva}}, \bibinfo {author} {\bibfnamefont {F.}~\bibnamefont {Calegari}}, \bibinfo {author} {\bibfnamefont {A.}~\bibnamefont {Palacios}}, \ and\ \bibinfo {author} {\bibfnamefont {F.}~\bibnamefont {Mart{\'\i}n}},\ }\href@noop {} {\bibfield  {journal} {\bibinfo  {journal} {Chemical reviews}\ }\textbf {\bibinfo {volume} {117}},\ \bibinfo {pages} {10760} (\bibinfo {year} {2017})}\BibitemShut {NoStop}%
\bibitem [{\citenamefont {Kowalewski}\ \emph {et~al.}(2017)\citenamefont {Kowalewski}, \citenamefont {Fingerhut}, \citenamefont {Dorfman}, \citenamefont {Bennett},\ and\ \citenamefont {Mukamel}}]{kowalewski2017}%
  \BibitemOpen
  \bibfield  {author} {\bibinfo {author} {\bibfnamefont {M.}~\bibnamefont {Kowalewski}}, \bibinfo {author} {\bibfnamefont {B.~P.}\ \bibnamefont {Fingerhut}}, \bibinfo {author} {\bibfnamefont {K.~E.}\ \bibnamefont {Dorfman}}, \bibinfo {author} {\bibfnamefont {K.}~\bibnamefont {Bennett}}, \ and\ \bibinfo {author} {\bibfnamefont {S.}~\bibnamefont {Mukamel}},\ }\href {\doibase 10.1021/acs.chemrev.7b00081} {\bibfield  {journal} {\bibinfo  {journal} {Chemical Reviews}\ }\textbf {\bibinfo {volume} {117}},\ \bibinfo {pages} {12165} (\bibinfo {year} {2017})}\BibitemShut {NoStop}%
\bibitem [{\citenamefont {Neville}\ \emph {et~al.}(2018)\citenamefont {Neville}, \citenamefont {Chergui}, \citenamefont {Stolow},\ and\ \citenamefont {Schuurman}}]{Neville2018}%
  \BibitemOpen
  \bibfield  {author} {\bibinfo {author} {\bibfnamefont {S.~P.}\ \bibnamefont {Neville}}, \bibinfo {author} {\bibfnamefont {M.}~\bibnamefont {Chergui}}, \bibinfo {author} {\bibfnamefont {A.}~\bibnamefont {Stolow}}, \ and\ \bibinfo {author} {\bibfnamefont {M.~S.}\ \bibnamefont {Schuurman}},\ }\href {\doibase 10.1103/PhysRevLett.120.243001} {\bibfield  {journal} {\bibinfo  {journal} {Physical review letters}\ }\textbf {\bibinfo {volume} {120}} (\bibinfo {year} {2018}),\ 10.1103/PhysRevLett.120.243001}\BibitemShut {NoStop}%
\bibitem [{\citenamefont {Li}\ \emph {et~al.}(2020)\citenamefont {Li}, \citenamefont {Govind}, \citenamefont {Isborn}, \citenamefont {DePrince},\ and\ \citenamefont {Lopata}}]{realtimemethods}%
  \BibitemOpen
  \bibfield  {author} {\bibinfo {author} {\bibfnamefont {X.}~\bibnamefont {Li}}, \bibinfo {author} {\bibfnamefont {N.}~\bibnamefont {Govind}}, \bibinfo {author} {\bibfnamefont {C.}~\bibnamefont {Isborn}}, \bibinfo {author} {\bibfnamefont {A.~E.~I.}\ \bibnamefont {DePrince}}, \ and\ \bibinfo {author} {\bibfnamefont {K.}~\bibnamefont {Lopata}},\ }\href {\doibase 10.1021/acs.chemrev.0c00223} {\bibfield  {journal} {\bibinfo  {journal} {Chemical Reviews}\ }\textbf {\bibinfo {volume} {120}},\ \bibinfo {pages} {9951} (\bibinfo {year} {2020})},\ \bibinfo {note} {pMID: 32813506},\ \Eprint {http://arxiv.org/abs/https://doi.org/10.1021/acs.chemrev.0c00223} {https://doi.org/10.1021/acs.chemrev.0c00223} \BibitemShut {NoStop}%
\bibitem [{\citenamefont {Zinchenko}\ \emph {et~al.}(2021)\citenamefont {Zinchenko}, \citenamefont {Ardana-Lamas}, \citenamefont {Seidu}, \citenamefont {Neville}, \citenamefont {van~der Veen}, \citenamefont {Lanfaloni}, \citenamefont {Schuurman},\ and\ \citenamefont {Worner}}]{Zinchenko2021}%
  \BibitemOpen
  \bibfield  {author} {\bibinfo {author} {\bibfnamefont {K.~S.}\ \bibnamefont {Zinchenko}}, \bibinfo {author} {\bibfnamefont {F.}~\bibnamefont {Ardana-Lamas}}, \bibinfo {author} {\bibfnamefont {I.}~\bibnamefont {Seidu}}, \bibinfo {author} {\bibfnamefont {S.~P.}\ \bibnamefont {Neville}}, \bibinfo {author} {\bibfnamefont {J.}~\bibnamefont {van~der Veen}}, \bibinfo {author} {\bibfnamefont {V.~U.}\ \bibnamefont {Lanfaloni}}, \bibinfo {author} {\bibfnamefont {M.~S.}\ \bibnamefont {Schuurman}}, \ and\ \bibinfo {author} {\bibfnamefont {H.~J.}\ \bibnamefont {Worner}},\ }\href {\doibase 10.1126/science.abf1656} {\bibfield  {journal} {\bibinfo  {journal} {Science}\ }\textbf {\bibinfo {volume} {371}},\ \bibinfo {pages} {489+} (\bibinfo {year} {2021})}\BibitemShut {NoStop}%
\bibitem [{\citenamefont {Cheng}\ \emph {et~al.}(2022)\citenamefont {Cheng}, \citenamefont {Singh}, \citenamefont {Matsika},\ and\ \citenamefont {Weinacht}}]{Cheng2022}%
  \BibitemOpen
  \bibfield  {author} {\bibinfo {author} {\bibfnamefont {C.}~\bibnamefont {Cheng}}, \bibinfo {author} {\bibfnamefont {V.}~\bibnamefont {Singh}}, \bibinfo {author} {\bibfnamefont {S.}~\bibnamefont {Matsika}}, \ and\ \bibinfo {author} {\bibfnamefont {T.}~\bibnamefont {Weinacht}},\ }\href {\doibase 10.1021/acs.jpca.2c04650} {\bibfield  {journal} {\bibinfo  {journal} {The Journal of Physical Chemistry A}\ } (\bibinfo {year} {2022}),\ 10.1021/acs.jpca.2c04650}\BibitemShut {NoStop}%
\bibitem [{\citenamefont {Freixas}\ \emph {et~al.}(2022)\citenamefont {Freixas}, \citenamefont {Keefer}, \citenamefont {Tretiak}, \citenamefont {Fernandez-Alberti},\ and\ \citenamefont {Mukamel}}]{Freixas2022}%
  \BibitemOpen
  \bibfield  {author} {\bibinfo {author} {\bibfnamefont {V.~M.}\ \bibnamefont {Freixas}}, \bibinfo {author} {\bibfnamefont {D.}~\bibnamefont {Keefer}}, \bibinfo {author} {\bibfnamefont {S.}~\bibnamefont {Tretiak}}, \bibinfo {author} {\bibfnamefont {S.}~\bibnamefont {Fernandez-Alberti}}, \ and\ \bibinfo {author} {\bibfnamefont {S.}~\bibnamefont {Mukamel}},\ }\href {\doibase 10.1039/d2sc00601d} {\bibfield  {journal} {\bibinfo  {journal} {Chemical Science}\ }\textbf {\bibinfo {volume} {13}},\ \bibinfo {pages} {6373} (\bibinfo {year} {2022})}\BibitemShut {NoStop}%
\bibitem [{\citenamefont {Centurion}, \citenamefont {Wolf},\ and\ \citenamefont {Yang}(2022)}]{wolf_ued}%
  \BibitemOpen
  \bibfield  {author} {\bibinfo {author} {\bibfnamefont {M.}~\bibnamefont {Centurion}}, \bibinfo {author} {\bibfnamefont {T.~J.}\ \bibnamefont {Wolf}}, \ and\ \bibinfo {author} {\bibfnamefont {J.}~\bibnamefont {Yang}},\ }\href {\doibase 10.1146/annurev-physchem-082720-010539} {\bibfield  {journal} {\bibinfo  {journal} {Annual Review of Physical Chemistry}\ }\textbf {\bibinfo {volume} {73}},\ \bibinfo {pages} {21} (\bibinfo {year} {2022})},\ \bibinfo {note} {pMID: 34724395},\ \Eprint {http://arxiv.org/abs/https://doi.org/10.1146/annurev-physchem-082720-010539} {https://doi.org/10.1146/annurev-physchem-082720-010539} \BibitemShut {NoStop}%
\bibitem [{\citenamefont {Williamson}\ \emph {et~al.}(1997)\citenamefont {Williamson}, \citenamefont {Cao}, \citenamefont {Ihee}, \citenamefont {Frey},\ and\ \citenamefont {Zewail}}]{Williamson1997}%
  \BibitemOpen
  \bibfield  {author} {\bibinfo {author} {\bibfnamefont {J.}~\bibnamefont {Williamson}}, \bibinfo {author} {\bibfnamefont {J.}~\bibnamefont {Cao}}, \bibinfo {author} {\bibfnamefont {H.}~\bibnamefont {Ihee}}, \bibinfo {author} {\bibfnamefont {H.}~\bibnamefont {Frey}}, \ and\ \bibinfo {author} {\bibfnamefont {A.}~\bibnamefont {Zewail}},\ }\href {\doibase 10.1038/386159a0} {\bibfield  {journal} {\bibinfo  {journal} {Nature}\ }\textbf {\bibinfo {volume} {386}},\ \bibinfo {pages} {159} (\bibinfo {year} {1997})}\BibitemShut {NoStop}%
\bibitem [{\citenamefont {Ihee}\ \emph {et~al.}(2001)\citenamefont {Ihee}, \citenamefont {Lobastov}, \citenamefont {Gomez}, \citenamefont {Goodson}, \citenamefont {Srinivasan}, \citenamefont {Ruan},\ and\ \citenamefont {Zewail}}]{Ihee2001}%
  \BibitemOpen
  \bibfield  {author} {\bibinfo {author} {\bibfnamefont {H.}~\bibnamefont {Ihee}}, \bibinfo {author} {\bibfnamefont {V.}~\bibnamefont {Lobastov}}, \bibinfo {author} {\bibfnamefont {U.}~\bibnamefont {Gomez}}, \bibinfo {author} {\bibfnamefont {B.}~\bibnamefont {Goodson}}, \bibinfo {author} {\bibfnamefont {R.}~\bibnamefont {Srinivasan}}, \bibinfo {author} {\bibfnamefont {C.}~\bibnamefont {Ruan}}, \ and\ \bibinfo {author} {\bibfnamefont {A.}~\bibnamefont {Zewail}},\ }\href {\doibase 10.1126/science.291.5503.458} {\bibfield  {journal} {\bibinfo  {journal} {Science}\ }\textbf {\bibinfo {volume} {291}},\ \bibinfo {pages} {458} (\bibinfo {year} {2001})}\BibitemShut {NoStop}%
\bibitem [{\citenamefont {Yang}\ \emph {et~al.}(2016{\natexlab{a}})\citenamefont {Yang}, \citenamefont {Guehr}, \citenamefont {Vecchione}, \citenamefont {Robinson}, \citenamefont {Li}, \citenamefont {Hartmann}, \citenamefont {Shen}, \citenamefont {Coffee}, \citenamefont {Corbett}, \citenamefont {Fry}, \citenamefont {Gaffney}, \citenamefont {Gorkhover}, \citenamefont {Hast}, \citenamefont {Jobe}, \citenamefont {Makasyuk}, \citenamefont {Reid}, \citenamefont {Robinson}, \citenamefont {Vetter}, \citenamefont {Wang}, \citenamefont {Weathersby}, \citenamefont {Yoneda}, \citenamefont {Centurion},\ and\ \citenamefont {Wang}}]{Yang2016a}%
  \BibitemOpen
  \bibfield  {author} {\bibinfo {author} {\bibfnamefont {J.}~\bibnamefont {Yang}}, \bibinfo {author} {\bibfnamefont {M.}~\bibnamefont {Guehr}}, \bibinfo {author} {\bibfnamefont {T.}~\bibnamefont {Vecchione}}, \bibinfo {author} {\bibfnamefont {M.~S.}\ \bibnamefont {Robinson}}, \bibinfo {author} {\bibfnamefont {R.}~\bibnamefont {Li}}, \bibinfo {author} {\bibfnamefont {N.}~\bibnamefont {Hartmann}}, \bibinfo {author} {\bibfnamefont {X.}~\bibnamefont {Shen}}, \bibinfo {author} {\bibfnamefont {R.}~\bibnamefont {Coffee}}, \bibinfo {author} {\bibfnamefont {J.}~\bibnamefont {Corbett}}, \bibinfo {author} {\bibfnamefont {A.}~\bibnamefont {Fry}}, \bibinfo {author} {\bibfnamefont {K.}~\bibnamefont {Gaffney}}, \bibinfo {author} {\bibfnamefont {T.}~\bibnamefont {Gorkhover}}, \bibinfo {author} {\bibfnamefont {C.}~\bibnamefont {Hast}}, \bibinfo {author} {\bibfnamefont {K.}~\bibnamefont {Jobe}}, \bibinfo {author} {\bibfnamefont {I.}~\bibnamefont {Makasyuk}}, \bibinfo {author} {\bibfnamefont {A.}~\bibnamefont {Reid}}, \bibinfo
  {author} {\bibfnamefont {J.}~\bibnamefont {Robinson}}, \bibinfo {author} {\bibfnamefont {S.}~\bibnamefont {Vetter}}, \bibinfo {author} {\bibfnamefont {F.}~\bibnamefont {Wang}}, \bibinfo {author} {\bibfnamefont {S.}~\bibnamefont {Weathersby}}, \bibinfo {author} {\bibfnamefont {C.}~\bibnamefont {Yoneda}}, \bibinfo {author} {\bibfnamefont {M.}~\bibnamefont {Centurion}}, \ and\ \bibinfo {author} {\bibfnamefont {X.}~\bibnamefont {Wang}},\ }\href {\doibase 10.1038/ncomms11232} {\bibfield  {journal} {\bibinfo  {journal} {Nature Communications}\ }\textbf {\bibinfo {volume} {7}} (\bibinfo {year} {2016}{\natexlab{a}}),\ 10.1038/ncomms11232}\BibitemShut {NoStop}%
\bibitem [{\citenamefont {Yang}\ \emph {et~al.}(2016{\natexlab{b}})\citenamefont {Yang}, \citenamefont {Guehr}, \citenamefont {Shen}, \citenamefont {Li}, \citenamefont {Vecchione}, \citenamefont {Coffee}, \citenamefont {Corbett}, \citenamefont {Fry}, \citenamefont {Hartmann}, \citenamefont {Hast}, \citenamefont {Hegazy}, \citenamefont {Jobe}, \citenamefont {Makasyuk}, \citenamefont {Robinson}, \citenamefont {Robinson}, \citenamefont {Vetter}, \citenamefont {Weathersby}, \citenamefont {Yoneda}, \citenamefont {Wang},\ and\ \citenamefont {Centurion}}]{Yang2016b}%
  \BibitemOpen
  \bibfield  {author} {\bibinfo {author} {\bibfnamefont {J.}~\bibnamefont {Yang}}, \bibinfo {author} {\bibfnamefont {M.}~\bibnamefont {Guehr}}, \bibinfo {author} {\bibfnamefont {X.}~\bibnamefont {Shen}}, \bibinfo {author} {\bibfnamefont {R.}~\bibnamefont {Li}}, \bibinfo {author} {\bibfnamefont {T.}~\bibnamefont {Vecchione}}, \bibinfo {author} {\bibfnamefont {R.}~\bibnamefont {Coffee}}, \bibinfo {author} {\bibfnamefont {J.}~\bibnamefont {Corbett}}, \bibinfo {author} {\bibfnamefont {A.}~\bibnamefont {Fry}}, \bibinfo {author} {\bibfnamefont {N.}~\bibnamefont {Hartmann}}, \bibinfo {author} {\bibfnamefont {C.}~\bibnamefont {Hast}}, \bibinfo {author} {\bibfnamefont {K.}~\bibnamefont {Hegazy}}, \bibinfo {author} {\bibfnamefont {K.}~\bibnamefont {Jobe}}, \bibinfo {author} {\bibfnamefont {I.}~\bibnamefont {Makasyuk}}, \bibinfo {author} {\bibfnamefont {J.}~\bibnamefont {Robinson}}, \bibinfo {author} {\bibfnamefont {M.~S.}\ \bibnamefont {Robinson}}, \bibinfo {author} {\bibfnamefont {S.}~\bibnamefont {Vetter}}, \bibinfo
  {author} {\bibfnamefont {S.}~\bibnamefont {Weathersby}}, \bibinfo {author} {\bibfnamefont {C.}~\bibnamefont {Yoneda}}, \bibinfo {author} {\bibfnamefont {X.}~\bibnamefont {Wang}}, \ and\ \bibinfo {author} {\bibfnamefont {M.}~\bibnamefont {Centurion}},\ }\href {\doibase 10.1103/PhysRevLett.117.153002} {\bibfield  {journal} {\bibinfo  {journal} {Physical review letters}\ }\textbf {\bibinfo {volume} {117}} (\bibinfo {year} {2016}{\natexlab{b}}),\ 10.1103/PhysRevLett.117.153002}\BibitemShut {NoStop}%
\bibitem [{\citenamefont {Yang}\ \emph {et~al.}(2018)\citenamefont {Yang}, \citenamefont {Zhu}, \citenamefont {Wolf}, \citenamefont {Li}, \citenamefont {Nunes}, \citenamefont {Coffee}, \citenamefont {Cryan}, \citenamefont {Guehr}, \citenamefont {Hegazy}, \citenamefont {Heinz}, \citenamefont {Jobe}, \citenamefont {Li}, \citenamefont {Shen}, \citenamefont {Veccione}, \citenamefont {Weathersby}, \citenamefont {Wilkin}, \citenamefont {Yoneda}, \citenamefont {Zheng}, \citenamefont {Martinez}, \citenamefont {Centurion},\ and\ \citenamefont {Wang}}]{Yang2018}%
  \BibitemOpen
  \bibfield  {author} {\bibinfo {author} {\bibfnamefont {J.}~\bibnamefont {Yang}}, \bibinfo {author} {\bibfnamefont {X.}~\bibnamefont {Zhu}}, \bibinfo {author} {\bibfnamefont {T.~J.~A.}\ \bibnamefont {Wolf}}, \bibinfo {author} {\bibfnamefont {Z.}~\bibnamefont {Li}}, \bibinfo {author} {\bibfnamefont {J.~P.~F.}\ \bibnamefont {Nunes}}, \bibinfo {author} {\bibfnamefont {R.}~\bibnamefont {Coffee}}, \bibinfo {author} {\bibfnamefont {J.~P.}\ \bibnamefont {Cryan}}, \bibinfo {author} {\bibfnamefont {M.}~\bibnamefont {Guehr}}, \bibinfo {author} {\bibfnamefont {K.}~\bibnamefont {Hegazy}}, \bibinfo {author} {\bibfnamefont {T.~F.}\ \bibnamefont {Heinz}}, \bibinfo {author} {\bibfnamefont {K.}~\bibnamefont {Jobe}}, \bibinfo {author} {\bibfnamefont {R.}~\bibnamefont {Li}}, \bibinfo {author} {\bibfnamefont {X.}~\bibnamefont {Shen}}, \bibinfo {author} {\bibfnamefont {T.}~\bibnamefont {Veccione}}, \bibinfo {author} {\bibfnamefont {S.}~\bibnamefont {Weathersby}}, \bibinfo {author} {\bibfnamefont {K.~J.}\ \bibnamefont {Wilkin}},
  \bibinfo {author} {\bibfnamefont {C.}~\bibnamefont {Yoneda}}, \bibinfo {author} {\bibfnamefont {Q.}~\bibnamefont {Zheng}}, \bibinfo {author} {\bibfnamefont {T.~J.}\ \bibnamefont {Martinez}}, \bibinfo {author} {\bibfnamefont {M.}~\bibnamefont {Centurion}}, \ and\ \bibinfo {author} {\bibfnamefont {X.}~\bibnamefont {Wang}},\ }\href {\doibase 10.1126/science.aat0049} {\bibfield  {journal} {\bibinfo  {journal} {Science}\ }\textbf {\bibinfo {volume} {361}},\ \bibinfo {pages} {64} (\bibinfo {year} {2018})}\BibitemShut {NoStop}%
\bibitem [{\citenamefont {Wolf}\ \emph {et~al.}(2019)\citenamefont {Wolf}, \citenamefont {Sanchez}, \citenamefont {Yang}, \citenamefont {Parrish}, \citenamefont {Nunes}, \citenamefont {Centurion}, \citenamefont {Coffee}, \citenamefont {Cryan}, \citenamefont {Guehr}, \citenamefont {Hegazy}, \citenamefont {Kirrander}, \citenamefont {Li}, \citenamefont {Ruddock}, \citenamefont {Shen}, \citenamefont {Vecchione}, \citenamefont {Weathersby}, \citenamefont {Weber}, \citenamefont {Wilkin}, \citenamefont {Yong}, \citenamefont {Zheng}, \citenamefont {Wang}, \citenamefont {Minitti},\ and\ \citenamefont {Martinez}}]{Wolf2019}%
  \BibitemOpen
  \bibfield  {author} {\bibinfo {author} {\bibfnamefont {T.~J.~A.}\ \bibnamefont {Wolf}}, \bibinfo {author} {\bibfnamefont {D.~M.}\ \bibnamefont {Sanchez}}, \bibinfo {author} {\bibfnamefont {J.}~\bibnamefont {Yang}}, \bibinfo {author} {\bibfnamefont {R.~M.}\ \bibnamefont {Parrish}}, \bibinfo {author} {\bibfnamefont {J.~P.~F.}\ \bibnamefont {Nunes}}, \bibinfo {author} {\bibfnamefont {M.}~\bibnamefont {Centurion}}, \bibinfo {author} {\bibfnamefont {R.}~\bibnamefont {Coffee}}, \bibinfo {author} {\bibfnamefont {J.~P.}\ \bibnamefont {Cryan}}, \bibinfo {author} {\bibfnamefont {M.}~\bibnamefont {Guehr}}, \bibinfo {author} {\bibfnamefont {K.}~\bibnamefont {Hegazy}}, \bibinfo {author} {\bibfnamefont {A.}~\bibnamefont {Kirrander}}, \bibinfo {author} {\bibfnamefont {R.~K.}\ \bibnamefont {Li}}, \bibinfo {author} {\bibfnamefont {J.}~\bibnamefont {Ruddock}}, \bibinfo {author} {\bibfnamefont {X.}~\bibnamefont {Shen}}, \bibinfo {author} {\bibfnamefont {T.}~\bibnamefont {Vecchione}}, \bibinfo {author} {\bibfnamefont {S.~P.}\
  \bibnamefont {Weathersby}}, \bibinfo {author} {\bibfnamefont {P.~M.}\ \bibnamefont {Weber}}, \bibinfo {author} {\bibfnamefont {K.}~\bibnamefont {Wilkin}}, \bibinfo {author} {\bibfnamefont {H.}~\bibnamefont {Yong}}, \bibinfo {author} {\bibfnamefont {Q.}~\bibnamefont {Zheng}}, \bibinfo {author} {\bibfnamefont {X.~J.}\ \bibnamefont {Wang}}, \bibinfo {author} {\bibfnamefont {M.~P.}\ \bibnamefont {Minitti}}, \ and\ \bibinfo {author} {\bibfnamefont {T.~J.}\ \bibnamefont {Martinez}},\ }\href {\doibase 10.1038/s41557-019-0252-7} {\bibfield  {journal} {\bibinfo  {journal} {Nature chemistry}\ }\textbf {\bibinfo {volume} {11}},\ \bibinfo {pages} {504} (\bibinfo {year} {2019})}\BibitemShut {NoStop}%
\bibitem [{\citenamefont {Stefanou}\ \emph {et~al.}(2017)\citenamefont {Stefanou}, \citenamefont {Saita}, \citenamefont {Shalashilin},\ and\ \citenamefont {Kirrander}}]{STEFANOU2017300}%
  \BibitemOpen
  \bibfield  {author} {\bibinfo {author} {\bibfnamefont {M.}~\bibnamefont {Stefanou}}, \bibinfo {author} {\bibfnamefont {K.}~\bibnamefont {Saita}}, \bibinfo {author} {\bibfnamefont {D.~V.}\ \bibnamefont {Shalashilin}}, \ and\ \bibinfo {author} {\bibfnamefont {A.}~\bibnamefont {Kirrander}},\ }\href {\doibase https://doi.org/10.1016/j.cplett.2017.03.007} {\bibfield  {journal} {\bibinfo  {journal} {Chemical Physics Letters}\ }\textbf {\bibinfo {volume} {683}},\ \bibinfo {pages} {300} (\bibinfo {year} {2017})},\ \bibinfo {note} {ahmed Zewail (1946-2016) Commemoration Issue of Chemical Physics Letters}\BibitemShut {NoStop}%
\bibitem [{\citenamefont {Parrish}\ and\ \citenamefont {Martínez}(2019)}]{parrish2019}%
  \BibitemOpen
  \bibfield  {author} {\bibinfo {author} {\bibfnamefont {R.~M.}\ \bibnamefont {Parrish}}\ and\ \bibinfo {author} {\bibfnamefont {T.~J.}\ \bibnamefont {Martínez}},\ }\href {\doibase 10.1021/acs.jctc.8b01051} {\bibfield  {journal} {\bibinfo  {journal} {Journal of Chemical Theory and Computation}\ }\textbf {\bibinfo {volume} {15}},\ \bibinfo {pages} {1523} (\bibinfo {year} {2019})},\ \bibinfo {note} {pMID: 30702882},\ \Eprint {http://arxiv.org/abs/https://doi.org/10.1021/acs.jctc.8b01051} {https://doi.org/10.1021/acs.jctc.8b01051} \BibitemShut {NoStop}%
\bibitem [{\citenamefont {Champenois}\ \emph {et~al.}(2021)\citenamefont {Champenois}, \citenamefont {Sanchez}, \citenamefont {Yang}, \citenamefont {Nunes}, \citenamefont {Attar}, \citenamefont {Centurion}, \citenamefont {Forbes}, \citenamefont {Gühr}, \citenamefont {Hegazy}, \citenamefont {Ji}, \citenamefont {Saha}, \citenamefont {Liu}, \citenamefont {Lin}, \citenamefont {Luo}, \citenamefont {Moore}, \citenamefont {Shen}, \citenamefont {Ware}, \citenamefont {Wang}, \citenamefont {Martínez},\ and\ \citenamefont {Wolf}}]{champenois2021aims}%
  \BibitemOpen
  \bibfield  {author} {\bibinfo {author} {\bibfnamefont {E.~G.}\ \bibnamefont {Champenois}}, \bibinfo {author} {\bibfnamefont {D.~M.}\ \bibnamefont {Sanchez}}, \bibinfo {author} {\bibfnamefont {J.}~\bibnamefont {Yang}}, \bibinfo {author} {\bibfnamefont {J.~P.~F.}\ \bibnamefont {Nunes}}, \bibinfo {author} {\bibfnamefont {A.}~\bibnamefont {Attar}}, \bibinfo {author} {\bibfnamefont {M.}~\bibnamefont {Centurion}}, \bibinfo {author} {\bibfnamefont {R.}~\bibnamefont {Forbes}}, \bibinfo {author} {\bibfnamefont {M.}~\bibnamefont {Gühr}}, \bibinfo {author} {\bibfnamefont {K.}~\bibnamefont {Hegazy}}, \bibinfo {author} {\bibfnamefont {F.}~\bibnamefont {Ji}}, \bibinfo {author} {\bibfnamefont {S.~K.}\ \bibnamefont {Saha}}, \bibinfo {author} {\bibfnamefont {Y.}~\bibnamefont {Liu}}, \bibinfo {author} {\bibfnamefont {M.-F.}\ \bibnamefont {Lin}}, \bibinfo {author} {\bibfnamefont {D.}~\bibnamefont {Luo}}, \bibinfo {author} {\bibfnamefont {B.}~\bibnamefont {Moore}}, \bibinfo {author} {\bibfnamefont {X.}~\bibnamefont {Shen}},
  \bibinfo {author} {\bibfnamefont {M.~R.}\ \bibnamefont {Ware}}, \bibinfo {author} {\bibfnamefont {X.~J.}\ \bibnamefont {Wang}}, \bibinfo {author} {\bibfnamefont {T.~J.}\ \bibnamefont {Martínez}}, \ and\ \bibinfo {author} {\bibfnamefont {T.~J.~A.}\ \bibnamefont {Wolf}},\ }\href {\doibase 10.1126/science.abk3132} {\bibfield  {journal} {\bibinfo  {journal} {Science}\ }\textbf {\bibinfo {volume} {374}},\ \bibinfo {pages} {178} (\bibinfo {year} {2021})},\ \Eprint {http://arxiv.org/abs/https://www.science.org/doi/pdf/10.1126/science.abk3132} {https://www.science.org/doi/pdf/10.1126/science.abk3132} \BibitemShut {NoStop}%
\bibitem [{\citenamefont {Liu}\ \emph {et~al.}(2023)\citenamefont {Liu}, \citenamefont {Sanchez}, \citenamefont {Ware}, \citenamefont {Champenois}, \citenamefont {Yang}, \citenamefont {Nunes}, \citenamefont {Attar}, \citenamefont {Centurion}, \citenamefont {Cryan}, \citenamefont {Forbes} \emph {et~al.}}]{liu2023rehybridization}%
  \BibitemOpen
  \bibfield  {author} {\bibinfo {author} {\bibfnamefont {Y.}~\bibnamefont {Liu}}, \bibinfo {author} {\bibfnamefont {D.~M.}\ \bibnamefont {Sanchez}}, \bibinfo {author} {\bibfnamefont {M.~R.}\ \bibnamefont {Ware}}, \bibinfo {author} {\bibfnamefont {E.~G.}\ \bibnamefont {Champenois}}, \bibinfo {author} {\bibfnamefont {J.}~\bibnamefont {Yang}}, \bibinfo {author} {\bibfnamefont {J.~P.~F.}\ \bibnamefont {Nunes}}, \bibinfo {author} {\bibfnamefont {A.}~\bibnamefont {Attar}}, \bibinfo {author} {\bibfnamefont {M.}~\bibnamefont {Centurion}}, \bibinfo {author} {\bibfnamefont {J.~P.}\ \bibnamefont {Cryan}}, \bibinfo {author} {\bibfnamefont {R.}~\bibnamefont {Forbes}},  \emph {et~al.},\ }\href@noop {} {\bibfield  {journal} {\bibinfo  {journal} {Nature Communications}\ }\textbf {\bibinfo {volume} {14}},\ \bibinfo {pages} {2795} (\bibinfo {year} {2023})}\BibitemShut {NoStop}%
\bibitem [{\citenamefont {Weathersby}\ \emph {et~al.}(2015)\citenamefont {Weathersby}, \citenamefont {Brown}, \citenamefont {Centurion}, \citenamefont {Chase}, \citenamefont {Coffee}, \citenamefont {Corbett}, \citenamefont {Eichner}, \citenamefont {Frisch}, \citenamefont {Fry}, \citenamefont {Gühr}, \citenamefont {Hartmann}, \citenamefont {Hast}, \citenamefont {Hettel}, \citenamefont {Jobe}, \citenamefont {Jongewaard}, \citenamefont {Lewandowski}, \citenamefont {Li}, \citenamefont {Lindenberg}, \citenamefont {Makasyuk}, \citenamefont {May}, \citenamefont {McCormick}, \citenamefont {Nguyen}, \citenamefont {Reid}, \citenamefont {Shen}, \citenamefont {Sokolowski-Tinten}, \citenamefont {Vecchione}, \citenamefont {Vetter}, \citenamefont {Wu}, \citenamefont {Yang}, \citenamefont {Dürr},\ and\ \citenamefont {Wang}}]{slac}%
  \BibitemOpen
  \bibfield  {author} {\bibinfo {author} {\bibfnamefont {S.~P.}\ \bibnamefont {Weathersby}}, \bibinfo {author} {\bibfnamefont {G.}~\bibnamefont {Brown}}, \bibinfo {author} {\bibfnamefont {M.}~\bibnamefont {Centurion}}, \bibinfo {author} {\bibfnamefont {T.~F.}\ \bibnamefont {Chase}}, \bibinfo {author} {\bibfnamefont {R.}~\bibnamefont {Coffee}}, \bibinfo {author} {\bibfnamefont {J.}~\bibnamefont {Corbett}}, \bibinfo {author} {\bibfnamefont {J.~P.}\ \bibnamefont {Eichner}}, \bibinfo {author} {\bibfnamefont {J.~C.}\ \bibnamefont {Frisch}}, \bibinfo {author} {\bibfnamefont {A.~R.}\ \bibnamefont {Fry}}, \bibinfo {author} {\bibfnamefont {M.}~\bibnamefont {Gühr}}, \bibinfo {author} {\bibfnamefont {N.}~\bibnamefont {Hartmann}}, \bibinfo {author} {\bibfnamefont {C.}~\bibnamefont {Hast}}, \bibinfo {author} {\bibfnamefont {R.}~\bibnamefont {Hettel}}, \bibinfo {author} {\bibfnamefont {R.~K.}\ \bibnamefont {Jobe}}, \bibinfo {author} {\bibfnamefont {E.~N.}\ \bibnamefont {Jongewaard}}, \bibinfo {author} {\bibfnamefont
  {J.~R.}\ \bibnamefont {Lewandowski}}, \bibinfo {author} {\bibfnamefont {R.~K.}\ \bibnamefont {Li}}, \bibinfo {author} {\bibfnamefont {A.~M.}\ \bibnamefont {Lindenberg}}, \bibinfo {author} {\bibfnamefont {I.}~\bibnamefont {Makasyuk}}, \bibinfo {author} {\bibfnamefont {J.~E.}\ \bibnamefont {May}}, \bibinfo {author} {\bibfnamefont {D.}~\bibnamefont {McCormick}}, \bibinfo {author} {\bibfnamefont {M.~N.}\ \bibnamefont {Nguyen}}, \bibinfo {author} {\bibfnamefont {A.~H.}\ \bibnamefont {Reid}}, \bibinfo {author} {\bibfnamefont {X.}~\bibnamefont {Shen}}, \bibinfo {author} {\bibfnamefont {K.}~\bibnamefont {Sokolowski-Tinten}}, \bibinfo {author} {\bibfnamefont {T.}~\bibnamefont {Vecchione}}, \bibinfo {author} {\bibfnamefont {S.~L.}\ \bibnamefont {Vetter}}, \bibinfo {author} {\bibfnamefont {J.}~\bibnamefont {Wu}}, \bibinfo {author} {\bibfnamefont {J.}~\bibnamefont {Yang}}, \bibinfo {author} {\bibfnamefont {H.~A.}\ \bibnamefont {Dürr}}, \ and\ \bibinfo {author} {\bibfnamefont {X.~J.}\ \bibnamefont {Wang}},\ }\href
  {\doibase 10.1063/1.4926994} {\bibfield  {journal} {\bibinfo  {journal} {Review of Scientific Instruments}\ }\textbf {\bibinfo {volume} {86}},\ \bibinfo {pages} {073702} (\bibinfo {year} {2015})},\ \Eprint {http://arxiv.org/abs/https://pubs.aip.org/aip/rsi/article-pdf/doi/10.1063/1.4926994/15949942/073702\_1\_online.pdf} {https://pubs.aip.org/aip/rsi/article-pdf/doi/10.1063/1.4926994/15949942/073702\_1\_online.pdf} \BibitemShut {NoStop}%
\bibitem [{Note1()}]{Note1}%
  \BibitemOpen
  \bibinfo {note} {This work was first submitted to arXiv on Jan. 15, 2024. https://doi.org/10.48550/arXiv.2401.08069}\BibitemShut {NoStop}%
\bibitem [{\citenamefont {Esch}\ and\ \citenamefont {Levine}(2020{\natexlab{a}})}]{esch2020a}%
  \BibitemOpen
  \bibfield  {author} {\bibinfo {author} {\bibfnamefont {M.~P.}\ \bibnamefont {Esch}}\ and\ \bibinfo {author} {\bibfnamefont {B.~G.}\ \bibnamefont {Levine}},\ }\href {\doibase 10.1063/5.0010081} {\bibfield  {journal} {\bibinfo  {journal} {The Journal of Chemical Physics}\ }\textbf {\bibinfo {volume} {152}} (\bibinfo {year} {2020}{\natexlab{a}}),\ 10.1063/5.0010081}\BibitemShut {NoStop}%
\bibitem [{\citenamefont {Esch}\ and\ \citenamefont {Levine}(2021)}]{tab21}%
  \BibitemOpen
  \bibfield  {author} {\bibinfo {author} {\bibfnamefont {M.~P.}\ \bibnamefont {Esch}}\ and\ \bibinfo {author} {\bibfnamefont {B.}~\bibnamefont {Levine}},\ }\href {\doibase 10.1063/5.0070686} {\bibfield  {journal} {\bibinfo  {journal} {The Journal of Chemical Physics}\ }\textbf {\bibinfo {volume} {155}},\ \bibinfo {pages} {214101} (\bibinfo {year} {2021})},\ \Eprint {http://arxiv.org/abs/https://pubs.aip.org/aip/jcp/article-pdf/doi/10.1063/5.0070686/14878146/214101\_1\_online.pdf} {https://pubs.aip.org/aip/jcp/article-pdf/doi/10.1063/5.0070686/14878146/214101\_1\_online.pdf} \BibitemShut {NoStop}%
\bibitem [{\citenamefont {Durden}\ \emph {et~al.}(2024)\citenamefont {Durden}, \citenamefont {Liang}, \citenamefont {Suchan}, \citenamefont {Teplukhin},\ and\ \citenamefont {Levine}}]{tabdms_zenodo}%
  \BibitemOpen
  \bibfield  {author} {\bibinfo {author} {\bibfnamefont {A.~S.}\ \bibnamefont {Durden}}, \bibinfo {author} {\bibfnamefont {F.}~\bibnamefont {Liang}}, \bibinfo {author} {\bibfnamefont {J.}~\bibnamefont {Suchan}}, \bibinfo {author} {\bibfnamefont {A.}~\bibnamefont {Teplukhin}}, \ and\ \bibinfo {author} {\bibfnamefont {B.~G.}\ \bibnamefont {Levine}},\ }\href {\doibase 10.5281/zenodo.10498050} {\  (\bibinfo {year} {2024}),\ 10.5281/zenodo.10498050}\BibitemShut {NoStop}%
\bibitem [{\citenamefont {Tully}(1990)}]{tully_errors}%
  \BibitemOpen
  \bibfield  {author} {\bibinfo {author} {\bibfnamefont {J.~C.}\ \bibnamefont {Tully}},\ }\href {\doibase 10.1063/1.459170} {\bibfield  {journal} {\bibinfo  {journal} {The Journal of Chemical Physics}\ }\textbf {\bibinfo {volume} {93}},\ \bibinfo {pages} {1061} (\bibinfo {year} {1990})},\ \Eprint {http://arxiv.org/abs/https://pubs.aip.org/aip/jcp/article-pdf/93/2/1061/11035518/1061\_1\_online.pdf} {https://pubs.aip.org/aip/jcp/article-pdf/93/2/1061/11035518/1061\_1\_online.pdf} \BibitemShut {NoStop}%
\bibitem [{\citenamefont {Bittner}\ and\ \citenamefont {Rossky}(1995)}]{bittnerrossky1}%
  \BibitemOpen
  \bibfield  {author} {\bibinfo {author} {\bibfnamefont {E.~R.}\ \bibnamefont {Bittner}}\ and\ \bibinfo {author} {\bibfnamefont {P.~J.}\ \bibnamefont {Rossky}},\ }\href {\doibase 10.1063/1.470177} {\bibfield  {journal} {\bibinfo  {journal} {The Journal of Chemical Physics}\ }\textbf {\bibinfo {volume} {103}},\ \bibinfo {pages} {8130} (\bibinfo {year} {1995})},\ \Eprint {http://arxiv.org/abs/https://pubs.aip.org/aip/jcp/article-pdf/103/18/8130/10778767/8130\_1\_online.pdf} {https://pubs.aip.org/aip/jcp/article-pdf/103/18/8130/10778767/8130\_1\_online.pdf} \BibitemShut {NoStop}%
\bibitem [{\citenamefont {Schwartz}\ \emph {et~al.}(1996)\citenamefont {Schwartz}, \citenamefont {Bittner}, \citenamefont {Prezhdo},\ and\ \citenamefont {Rossky}}]{bittnerrossky2}%
  \BibitemOpen
  \bibfield  {author} {\bibinfo {author} {\bibfnamefont {B.~J.}\ \bibnamefont {Schwartz}}, \bibinfo {author} {\bibfnamefont {E.~R.}\ \bibnamefont {Bittner}}, \bibinfo {author} {\bibfnamefont {O.~V.}\ \bibnamefont {Prezhdo}}, \ and\ \bibinfo {author} {\bibfnamefont {P.~J.}\ \bibnamefont {Rossky}},\ }\href {\doibase 10.1063/1.471326} {\bibfield  {journal} {\bibinfo  {journal} {The Journal of Chemical Physics}\ }\textbf {\bibinfo {volume} {104}},\ \bibinfo {pages} {5942} (\bibinfo {year} {1996})},\ \Eprint {http://arxiv.org/abs/https://pubs.aip.org/aip/jcp/article-pdf/104/15/5942/8105629/5942\_1\_online.pdf} {https://pubs.aip.org/aip/jcp/article-pdf/104/15/5942/8105629/5942\_1\_online.pdf} \BibitemShut {NoStop}%
\bibitem [{\citenamefont {Esch}, \citenamefont {Shu},\ and\ \citenamefont {Levine}(2019)}]{Esch2019}%
  \BibitemOpen
  \bibfield  {author} {\bibinfo {author} {\bibfnamefont {M.~P.}\ \bibnamefont {Esch}}, \bibinfo {author} {\bibfnamefont {Y.}~\bibnamefont {Shu}}, \ and\ \bibinfo {author} {\bibfnamefont {B.~G.}\ \bibnamefont {Levine}},\ }\href {\doibase 10.1021/acs.jpca.9b00952} {\bibfield  {journal} {\bibinfo  {journal} {The Journal of Physical Chemistry A}\ }\textbf {\bibinfo {volume} {123}},\ \bibinfo {pages} {2661} (\bibinfo {year} {2019})}\BibitemShut {NoStop}%
\bibitem [{\citenamefont {Esch}\ and\ \citenamefont {Levine}(2020{\natexlab{b}})}]{esch2020b}%
  \BibitemOpen
  \bibfield  {author} {\bibinfo {author} {\bibfnamefont {M.~P.}\ \bibnamefont {Esch}}\ and\ \bibinfo {author} {\bibfnamefont {B.~G.}\ \bibnamefont {Levine}},\ }\href {\doibase 10.1063/5.0022529} {\bibfield  {journal} {\bibinfo  {journal} {The Journal of Chemical Physics}\ }\textbf {\bibinfo {volume} {153}} (\bibinfo {year} {2020}{\natexlab{b}}),\ 10.1063/5.0022529}\BibitemShut {NoStop}%
\bibitem [{\citenamefont {Peng}, \citenamefont {Fales},\ and\ \citenamefont {Levine}(2018)}]{tdci}%
  \BibitemOpen
  \bibfield  {author} {\bibinfo {author} {\bibfnamefont {W.-T.}\ \bibnamefont {Peng}}, \bibinfo {author} {\bibfnamefont {B.~S.}\ \bibnamefont {Fales}}, \ and\ \bibinfo {author} {\bibfnamefont {B.~G.}\ \bibnamefont {Levine}},\ }\href {\doibase 10.1021/acs.jctc.8b00381} {\bibfield  {journal} {\bibinfo  {journal} {Journal of Chemical Theory and Computation}\ }\textbf {\bibinfo {volume} {14}},\ \bibinfo {pages} {4129} (\bibinfo {year} {2018})},\ \bibinfo {note} {pMID: 29986143},\ \Eprint {http://arxiv.org/abs/https://doi.org/10.1021/acs.jctc.8b00381} {https://doi.org/10.1021/acs.jctc.8b00381} \BibitemShut {NoStop}%
\bibitem [{\citenamefont {Durden}\ and\ \citenamefont {Levine}(2022)}]{durden2022}%
  \BibitemOpen
  \bibfield  {author} {\bibinfo {author} {\bibfnamefont {A.~S.}\ \bibnamefont {Durden}}\ and\ \bibinfo {author} {\bibfnamefont {B.~G.}\ \bibnamefont {Levine}},\ }\href {\doibase 10.1021/acs.jctc.1c01009} {\bibfield  {journal} {\bibinfo  {journal} {Journal of Chemical Theory and Computation}\ }\textbf {\bibinfo {volume} {18}},\ \bibinfo {pages} {795} (\bibinfo {year} {2022})}\BibitemShut {NoStop}%
\bibitem [{\citenamefont {Gray}\ and\ \citenamefont {Manolopoulos}(1996)}]{gray1996}%
  \BibitemOpen
  \bibfield  {author} {\bibinfo {author} {\bibfnamefont {S.}~\bibnamefont {Gray}}\ and\ \bibinfo {author} {\bibfnamefont {D.}~\bibnamefont {Manolopoulos}},\ }\href {\doibase 10.1063/1.471428} {\bibfield  {journal} {\bibinfo  {journal} {The Journal of Chemical Physics}\ }\textbf {\bibinfo {volume} {104}},\ \bibinfo {pages} {7099} (\bibinfo {year} {1996})}\BibitemShut {NoStop}%
\bibitem [{\citenamefont {Seritan}\ \emph {et~al.}(2021)\citenamefont {Seritan}, \citenamefont {Bannwarth}, \citenamefont {Fales}, \citenamefont {Hohenstein}, \citenamefont {Isborn}, \citenamefont {Kokkila-Schumacher}, \citenamefont {Li}, \citenamefont {Liu}, \citenamefont {Luehr}, \citenamefont {Snyder~Jr.}, \citenamefont {Song}, \citenamefont {Titov}, \citenamefont {Ufimtsev}, \citenamefont {Wang},\ and\ \citenamefont {Martínez}}]{terachem21}%
  \BibitemOpen
  \bibfield  {author} {\bibinfo {author} {\bibfnamefont {S.}~\bibnamefont {Seritan}}, \bibinfo {author} {\bibfnamefont {C.}~\bibnamefont {Bannwarth}}, \bibinfo {author} {\bibfnamefont {B.~S.}\ \bibnamefont {Fales}}, \bibinfo {author} {\bibfnamefont {E.~G.}\ \bibnamefont {Hohenstein}}, \bibinfo {author} {\bibfnamefont {C.~M.}\ \bibnamefont {Isborn}}, \bibinfo {author} {\bibfnamefont {S.~I.~L.}\ \bibnamefont {Kokkila-Schumacher}}, \bibinfo {author} {\bibfnamefont {X.}~\bibnamefont {Li}}, \bibinfo {author} {\bibfnamefont {F.}~\bibnamefont {Liu}}, \bibinfo {author} {\bibfnamefont {N.}~\bibnamefont {Luehr}}, \bibinfo {author} {\bibfnamefont {J.~W.}\ \bibnamefont {Snyder~Jr.}}, \bibinfo {author} {\bibfnamefont {C.}~\bibnamefont {Song}}, \bibinfo {author} {\bibfnamefont {A.~V.}\ \bibnamefont {Titov}}, \bibinfo {author} {\bibfnamefont {I.~S.}\ \bibnamefont {Ufimtsev}}, \bibinfo {author} {\bibfnamefont {L.-P.}\ \bibnamefont {Wang}}, \ and\ \bibinfo {author} {\bibfnamefont {T.~J.}\ \bibnamefont {Martínez}},\ }\href
  {\doibase https://doi.org/10.1002/wcms.1494} {\bibfield  {journal} {\bibinfo  {journal} {WIREs Computational Molecular Science}\ }\textbf {\bibinfo {volume} {11}},\ \bibinfo {pages} {e1494} (\bibinfo {year} {2021})},\ \Eprint {http://arxiv.org/abs/https://wires.onlinelibrary.wiley.com/doi/pdf/10.1002/wcms.1494} {https://wires.onlinelibrary.wiley.com/doi/pdf/10.1002/wcms.1494} \BibitemShut {NoStop}%
\bibitem [{\citenamefont {Ufimtsev}\ and\ \citenamefont {Martinez}(2008)}]{Ufimtsev2008}%
  \BibitemOpen
  \bibfield  {author} {\bibinfo {author} {\bibfnamefont {I.~S.}\ \bibnamefont {Ufimtsev}}\ and\ \bibinfo {author} {\bibfnamefont {T.~J.}\ \bibnamefont {Martinez}},\ }\href {\doibase 10.1021/ct700268q} {\bibfield  {journal} {\bibinfo  {journal} {Journal of Chemical Theory and Computation}\ }\textbf {\bibinfo {volume} {4}},\ \bibinfo {pages} {222} (\bibinfo {year} {2008})}\BibitemShut {NoStop}%
\bibitem [{\citenamefont {Ufimtsev}\ and\ \citenamefont {Martinez}(2009{\natexlab{a}})}]{Ufimtsev2009a}%
  \BibitemOpen
  \bibfield  {author} {\bibinfo {author} {\bibfnamefont {I.~S.}\ \bibnamefont {Ufimtsev}}\ and\ \bibinfo {author} {\bibfnamefont {T.~J.}\ \bibnamefont {Martinez}},\ }\href {\doibase 10.1021/ct800526s} {\bibfield  {journal} {\bibinfo  {journal} {Journal of Chemical Theory and Computation}\ }\textbf {\bibinfo {volume} {5}},\ \bibinfo {pages} {1004} (\bibinfo {year} {2009}{\natexlab{a}})}\BibitemShut {NoStop}%
\bibitem [{\citenamefont {Ufimtsev}\ and\ \citenamefont {Martinez}(2009{\natexlab{b}})}]{Ufimtsev2009b}%
  \BibitemOpen
  \bibfield  {author} {\bibinfo {author} {\bibfnamefont {I.~S.}\ \bibnamefont {Ufimtsev}}\ and\ \bibinfo {author} {\bibfnamefont {T.~J.}\ \bibnamefont {Martinez}},\ }\href {\doibase 10.1021/ct9003004} {\bibfield  {journal} {\bibinfo  {journal} {Journal of Chemical Theory and Computation}\ }\textbf {\bibinfo {volume} {5}},\ \bibinfo {pages} {2619} (\bibinfo {year} {2009}{\natexlab{b}})}\BibitemShut {NoStop}%
\bibitem [{\citenamefont {Fales}\ and\ \citenamefont {Levine}(2015)}]{Fales2015}%
  \BibitemOpen
  \bibfield  {author} {\bibinfo {author} {\bibfnamefont {B.~S.}\ \bibnamefont {Fales}}\ and\ \bibinfo {author} {\bibfnamefont {B.~G.}\ \bibnamefont {Levine}},\ }\href {\doibase 10.1021/acs.jctc.5b00634} {\bibfield  {journal} {\bibinfo  {journal} {Journal of Chemical Theory and Computation}\ }\textbf {\bibinfo {volume} {11}},\ \bibinfo {pages} {4708} (\bibinfo {year} {2015})}\BibitemShut {NoStop}%
\bibitem [{\citenamefont {Hohenstein}\ \emph {et~al.}(2015)\citenamefont {Hohenstein}, \citenamefont {Bouduban}, \citenamefont {Song}, \citenamefont {Luehr}, \citenamefont {Ufimtsev},\ and\ \citenamefont {Martinez}}]{Hohenstein2015}%
  \BibitemOpen
  \bibfield  {author} {\bibinfo {author} {\bibfnamefont {E.~G.}\ \bibnamefont {Hohenstein}}, \bibinfo {author} {\bibfnamefont {M.~E.~F.}\ \bibnamefont {Bouduban}}, \bibinfo {author} {\bibfnamefont {C.}~\bibnamefont {Song}}, \bibinfo {author} {\bibfnamefont {N.}~\bibnamefont {Luehr}}, \bibinfo {author} {\bibfnamefont {I.~S.}\ \bibnamefont {Ufimtsev}}, \ and\ \bibinfo {author} {\bibfnamefont {T.~J.}\ \bibnamefont {Martinez}},\ }\href {\doibase 10.1063/1.4923259} {\bibfield  {journal} {\bibinfo  {journal} {The Journal of Chemical Physics}\ }\textbf {\bibinfo {volume} {143}} (\bibinfo {year} {2015}),\ 10.1063/1.4923259}\BibitemShut {NoStop}%
\bibitem [{\citenamefont {Liu}\ and\ \citenamefont {Fang}(2016)}]{prevaims}%
  \BibitemOpen
  \bibfield  {author} {\bibinfo {author} {\bibfnamefont {L.}~\bibnamefont {Liu}}\ and\ \bibinfo {author} {\bibfnamefont {W.-H.}\ \bibnamefont {Fang}},\ }\href {\doibase 10.1063/1.4945782} {\bibfield  {journal} {\bibinfo  {journal} {The Journal of Chemical Physics}\ }\textbf {\bibinfo {volume} {144}},\ \bibinfo {pages} {144317} (\bibinfo {year} {2016})},\ \Eprint {http://arxiv.org/abs/https://pubs.aip.org/aip/jcp/article-pdf/doi/10.1063/1.4945782/13555376/144317\_1\_online.pdf} {https://pubs.aip.org/aip/jcp/article-pdf/doi/10.1063/1.4945782/13555376/144317\_1\_online.pdf} \BibitemShut {NoStop}%
\bibitem [{\citenamefont {Xia}\ \emph {et~al.}(2015)\citenamefont {Xia}, \citenamefont {Liu}, \citenamefont {Fang},\ and\ \citenamefont {Cui}}]{prevringop}%
  \BibitemOpen
  \bibfield  {author} {\bibinfo {author} {\bibfnamefont {S.-H.}\ \bibnamefont {Xia}}, \bibinfo {author} {\bibfnamefont {X.-Y.}\ \bibnamefont {Liu}}, \bibinfo {author} {\bibfnamefont {Q.}~\bibnamefont {Fang}}, \ and\ \bibinfo {author} {\bibfnamefont {G.}~\bibnamefont {Cui}},\ }\href {\doibase 10.1021/acs.jpca.5b00302} {\bibfield  {journal} {\bibinfo  {journal} {The Journal of Physical Chemistry A}\ }\textbf {\bibinfo {volume} {119}},\ \bibinfo {pages} {3569} (\bibinfo {year} {2015})},\ \bibinfo {note} {pMID: 25807113},\ \Eprint {http://arxiv.org/abs/https://doi.org/10.1021/acs.jpca.5b00302} {https://doi.org/10.1021/acs.jpca.5b00302} \BibitemShut {NoStop}%
\bibitem [{\citenamefont {Diau}, \citenamefont {Kötting},\ and\ \citenamefont {Zewail}(2001)}]{prevnorish}%
  \BibitemOpen
  \bibfield  {author} {\bibinfo {author} {\bibfnamefont {E.~W.-G.}\ \bibnamefont {Diau}}, \bibinfo {author} {\bibfnamefont {C.}~\bibnamefont {Kötting}}, \ and\ \bibinfo {author} {\bibfnamefont {A.~H.}\ \bibnamefont {Zewail}},\ }\href {\doibase https://doi.org/10.1002/1439-7641(20010518)2:5<294::AID-CPHC294>3.0.CO;2-5} {\bibfield  {journal} {\bibinfo  {journal} {ChemPhysChem}\ }\textbf {\bibinfo {volume} {2}},\ \bibinfo {pages} {294} (\bibinfo {year} {2001})}\BibitemShut {NoStop}%
\bibitem [{\citenamefont {Hemminger}\ and\ \citenamefont {Lee}(2003)}]{isctime}%
  \BibitemOpen
  \bibfield  {author} {\bibinfo {author} {\bibfnamefont {J.~C.}\ \bibnamefont {Hemminger}}\ and\ \bibinfo {author} {\bibfnamefont {E.~K.~C.}\ \bibnamefont {Lee}},\ }\href {\doibase 10.1063/1.1677033} {\bibfield  {journal} {\bibinfo  {journal} {The Journal of Chemical Physics}\ }\textbf {\bibinfo {volume} {56}},\ \bibinfo {pages} {5284} (\bibinfo {year} {2003})},\ \Eprint {http://arxiv.org/abs/https://pubs.aip.org/aip/jcp/article-pdf/56/11/5284/11101722/5284\_1\_online.pdf} {https://pubs.aip.org/aip/jcp/article-pdf/56/11/5284/11101722/5284\_1\_online.pdf} \BibitemShut {NoStop}%
\bibitem [{\citenamefont {Udvarhazi}\ and\ \citenamefont {El‐Sayed}(2004)}]{elsayed_ryd}%
  \BibitemOpen
  \bibfield  {author} {\bibinfo {author} {\bibfnamefont {A.}~\bibnamefont {Udvarhazi}}\ and\ \bibinfo {author} {\bibfnamefont {M.~A.}\ \bibnamefont {El‐Sayed}},\ }\href {\doibase 10.1063/1.1696427} {\bibfield  {journal} {\bibinfo  {journal} {The Journal of Chemical Physics}\ }\textbf {\bibinfo {volume} {42}},\ \bibinfo {pages} {3335} (\bibinfo {year} {2004})},\ \Eprint {http://arxiv.org/abs/https://pubs.aip.org/aip/jcp/article-pdf/42/9/3335/11206722/3335\_1\_online.pdf} {https://pubs.aip.org/aip/jcp/article-pdf/42/9/3335/11206722/3335\_1\_online.pdf} \BibitemShut {NoStop}%
\bibitem [{\citenamefont {Hemminger}, \citenamefont {Carless},\ and\ \citenamefont {Lee}(1973)}]{leeval}%
  \BibitemOpen
  \bibfield  {author} {\bibinfo {author} {\bibfnamefont {J.}~\bibnamefont {Hemminger}}, \bibinfo {author} {\bibfnamefont {H.~A.}\ \bibnamefont {Carless}}, \ and\ \bibinfo {author} {\bibfnamefont {E.~K.}\ \bibnamefont {Lee}},\ }\href@noop {} {\bibfield  {journal} {\bibinfo  {journal} {Journal of the American Chemical Society}\ }\textbf {\bibinfo {volume} {95}},\ \bibinfo {pages} {682} (\bibinfo {year} {1973})}\BibitemShut {NoStop}%
\bibitem [{\citenamefont {Levine}\ \emph {et~al.}(2021)\citenamefont {Levine}, \citenamefont {Durden}, \citenamefont {Esch}, \citenamefont {Liang},\ and\ \citenamefont {Shu}}]{Levine2021}%
  \BibitemOpen
  \bibfield  {author} {\bibinfo {author} {\bibfnamefont {B.~G.}\ \bibnamefont {Levine}}, \bibinfo {author} {\bibfnamefont {A.~S.}\ \bibnamefont {Durden}}, \bibinfo {author} {\bibfnamefont {M.~P.}\ \bibnamefont {Esch}}, \bibinfo {author} {\bibfnamefont {F.}~\bibnamefont {Liang}}, \ and\ \bibinfo {author} {\bibfnamefont {Y.}~\bibnamefont {Shu}},\ }\href {\doibase 10.1063/5.0042147} {\bibfield  {journal} {\bibinfo  {journal} {The Journal of Chemical Physics}\ }\textbf {\bibinfo {volume} {154}},\ \bibinfo {pages} {090902} (\bibinfo {year} {2021})},\ \Eprint {http://arxiv.org/abs/https://doi.org/10.1063/5.0042147} {https://doi.org/10.1063/5.0042147} \BibitemShut {NoStop}%
\bibitem [{\citenamefont {Slavíček}\ and\ \citenamefont {Martínez}(2010)}]{fomo}%
  \BibitemOpen
  \bibfield  {author} {\bibinfo {author} {\bibfnamefont {P.}~\bibnamefont {Slavíček}}\ and\ \bibinfo {author} {\bibfnamefont {T.~J.}\ \bibnamefont {Martínez}},\ }\href {\doibase 10.1063/1.3436501} {\bibfield  {journal} {\bibinfo  {journal} {The Journal of Chemical Physics}\ }\textbf {\bibinfo {volume} {132}},\ \bibinfo {pages} {234102} (\bibinfo {year} {2010})},\ \Eprint {http://arxiv.org/abs/https://pubs.aip.org/aip/jcp/article-pdf/doi/10.1063/1.3436501/16031951/234102\_1\_online.pdf} {https://pubs.aip.org/aip/jcp/article-pdf/doi/10.1063/1.3436501/16031951/234102\_1\_online.pdf} \BibitemShut {NoStop}%
\bibitem [{\citenamefont {Granucci}, \citenamefont {Persico},\ and\ \citenamefont {Toniolo}(2001)}]{granucci2001}%
  \BibitemOpen
  \bibfield  {author} {\bibinfo {author} {\bibfnamefont {G.}~\bibnamefont {Granucci}}, \bibinfo {author} {\bibfnamefont {M.}~\bibnamefont {Persico}}, \ and\ \bibinfo {author} {\bibfnamefont {A.}~\bibnamefont {Toniolo}},\ }\href {\doibase 10.1063/1.1376633} {\bibfield  {journal} {\bibinfo  {journal} {The Journal of Chemical Physics}\ }\textbf {\bibinfo {volume} {114}},\ \bibinfo {pages} {10608} (\bibinfo {year} {2001})}\BibitemShut {NoStop}%
\bibitem [{\citenamefont {Salvat}, \citenamefont {Jablonski},\ and\ \citenamefont {Powell}(2005)}]{elsepa}%
  \BibitemOpen
  \bibfield  {author} {\bibinfo {author} {\bibfnamefont {F.}~\bibnamefont {Salvat}}, \bibinfo {author} {\bibfnamefont {A.}~\bibnamefont {Jablonski}}, \ and\ \bibinfo {author} {\bibfnamefont {C.~J.}\ \bibnamefont {Powell}},\ }\href {\doibase https://doi.org/10.1016/j.cpc.2004.09.006} {\bibfield  {journal} {\bibinfo  {journal} {Computer Physics Communications}\ }\textbf {\bibinfo {volume} {165}},\ \bibinfo {pages} {157} (\bibinfo {year} {2005})}\BibitemShut {NoStop}%
\bibitem [{\citenamefont {Denschlag}\ and\ \citenamefont {Lee}(1968)}]{denschlag1968benzene}%
  \BibitemOpen
  \bibfield  {author} {\bibinfo {author} {\bibfnamefont {H.}~\bibnamefont {Denschlag}}\ and\ \bibinfo {author} {\bibfnamefont {E.~K.}\ \bibnamefont {Lee}},\ }\href@noop {} {\bibfield  {journal} {\bibinfo  {journal} {Journal of the American Chemical Society}\ }\textbf {\bibinfo {volume} {90}},\ \bibinfo {pages} {3628} (\bibinfo {year} {1968})}\BibitemShut {NoStop}%
\bibitem [{\citenamefont {Lee}\ and\ \citenamefont {Lee}(2003)}]{cycbut_productratio_exp}%
  \BibitemOpen
  \bibfield  {author} {\bibinfo {author} {\bibfnamefont {N.~E.}\ \bibnamefont {Lee}}\ and\ \bibinfo {author} {\bibfnamefont {E.~K.~C.}\ \bibnamefont {Lee}},\ }\href {\doibase 10.1063/1.1671339} {\bibfield  {journal} {\bibinfo  {journal} {The Journal of Chemical Physics}\ }\textbf {\bibinfo {volume} {50}},\ \bibinfo {pages} {2094} (\bibinfo {year} {2003})},\ \Eprint {http://arxiv.org/abs/https://pubs.aip.org/aip/jcp/article-pdf/50/5/2094/11098793/2094\_1\_online.pdf} {https://pubs.aip.org/aip/jcp/article-pdf/50/5/2094/11098793/2094\_1\_online.pdf} \BibitemShut {NoStop}%
\bibitem [{\citenamefont {Lee}(1977)}]{laserphotolee}%
  \BibitemOpen
  \bibfield  {author} {\bibinfo {author} {\bibfnamefont {E.~K.~C.}\ \bibnamefont {Lee}},\ }\href {\doibase 10.1021/ar50117a002} {\bibfield  {journal} {\bibinfo  {journal} {Accounts of Chemical Research}\ }\textbf {\bibinfo {volume} {10}},\ \bibinfo {pages} {319} (\bibinfo {year} {1977})},\ \Eprint {http://arxiv.org/abs/https://doi.org/10.1021/ar50117a002} {https://doi.org/10.1021/ar50117a002} \BibitemShut {NoStop}%
\bibitem [{\citenamefont {Heller}(1981)}]{heller1981semiclassical}%
  \BibitemOpen
  \bibfield  {author} {\bibinfo {author} {\bibfnamefont {E.~J.}\ \bibnamefont {Heller}},\ }\href@noop {} {\bibfield  {journal} {\bibinfo  {journal} {Accounts of Chemical Research}\ }\textbf {\bibinfo {volume} {14}},\ \bibinfo {pages} {368} (\bibinfo {year} {1981})}\BibitemShut {NoStop}%
\bibitem [{\citenamefont {Kuhlman}, \citenamefont {Solling},\ and\ \citenamefont {Moller}(2012)}]{Kuhlman2012}%
  \BibitemOpen
  \bibfield  {author} {\bibinfo {author} {\bibfnamefont {T.~S.}\ \bibnamefont {Kuhlman}}, \bibinfo {author} {\bibfnamefont {T.~I.}\ \bibnamefont {Solling}}, \ and\ \bibinfo {author} {\bibfnamefont {K.~B.}\ \bibnamefont {Moller}},\ }\href {\doibase 10.1002/cphc.201100929} {\bibfield  {journal} {\bibinfo  {journal} {CHEMPHYSCHEM}\ }\textbf {\bibinfo {volume} {13}},\ \bibinfo {pages} {820} (\bibinfo {year} {2012})}\BibitemShut {NoStop}%
\bibitem [{\citenamefont {Marcus}(2004)}]{rrkm}%
  \BibitemOpen
  \bibfield  {author} {\bibinfo {author} {\bibfnamefont {R.~A.}\ \bibnamefont {Marcus}},\ }\href {\doibase 10.1063/1.1700424} {\bibfield  {journal} {\bibinfo  {journal} {The Journal of Chemical Physics}\ }\textbf {\bibinfo {volume} {20}},\ \bibinfo {pages} {359} (\bibinfo {year} {2004})},\ \Eprint {http://arxiv.org/abs/https://pubs.aip.org/aip/jcp/article-pdf/20/3/359/8113340/359\_1\_online.pdf} {https://pubs.aip.org/aip/jcp/article-pdf/20/3/359/8113340/359\_1\_online.pdf} \BibitemShut {NoStop}%
\bibitem [{\citenamefont {Beyer}\ and\ \citenamefont {Swinehart}(1973)}]{bsa}%
  \BibitemOpen
  \bibfield  {author} {\bibinfo {author} {\bibfnamefont {T.}~\bibnamefont {Beyer}}\ and\ \bibinfo {author} {\bibfnamefont {D.~F.}\ \bibnamefont {Swinehart}},\ }\href {\doibase 10.1145/362248.362275} {\ \textbf {\bibinfo {volume} {16}} (\bibinfo {year} {1973}),\ 10.1145/362248.362275}\BibitemShut {NoStop}%
\bibitem [{\citenamefont {Kao}\ \emph {et~al.}(2020)\citenamefont {Kao}, \citenamefont {Venkatraman}, \citenamefont {Ashfold},\ and\ \citenamefont {Orr-Ewing}}]{prevTAS}%
  \BibitemOpen
  \bibfield  {author} {\bibinfo {author} {\bibfnamefont {M.-H.}\ \bibnamefont {Kao}}, \bibinfo {author} {\bibfnamefont {R.~K.}\ \bibnamefont {Venkatraman}}, \bibinfo {author} {\bibfnamefont {M.~N.~R.}\ \bibnamefont {Ashfold}}, \ and\ \bibinfo {author} {\bibfnamefont {A.~J.}\ \bibnamefont {Orr-Ewing}},\ }\href {\doibase 10.1039/C9SC05208A} {\bibfield  {journal} {\bibinfo  {journal} {Chem. Sci.}\ }\textbf {\bibinfo {volume} {11}},\ \bibinfo {pages} {1991} (\bibinfo {year} {2020})}\BibitemShut {NoStop}%
\bibitem [{\citenamefont {O'Toole}\ \emph {et~al.}(1991)\citenamefont {O'Toole}, \citenamefont {Brint}, \citenamefont {Kosmidis}, \citenamefont {Boulakis},\ and\ \citenamefont {Tsekeris}}]{expspec}%
  \BibitemOpen
  \bibfield  {author} {\bibinfo {author} {\bibfnamefont {L.}~\bibnamefont {O'Toole}}, \bibinfo {author} {\bibfnamefont {P.}~\bibnamefont {Brint}}, \bibinfo {author} {\bibfnamefont {C.}~\bibnamefont {Kosmidis}}, \bibinfo {author} {\bibfnamefont {G.}~\bibnamefont {Boulakis}}, \ and\ \bibinfo {author} {\bibfnamefont {P.}~\bibnamefont {Tsekeris}},\ }\href {\doibase 10.1039/FT9918703343} {\bibfield  {journal} {\bibinfo  {journal} {J. Chem. Soc., Faraday Trans.}\ }\textbf {\bibinfo {volume} {87}},\ \bibinfo {pages} {3343} (\bibinfo {year} {1991})}\BibitemShut {NoStop}%
\bibitem [{\citenamefont {Thompson}, \citenamefont {Punwong},\ and\ \citenamefont {Martínez}(2010)}]{alphadecoh}%
  \BibitemOpen
  \bibfield  {author} {\bibinfo {author} {\bibfnamefont {A.~L.}\ \bibnamefont {Thompson}}, \bibinfo {author} {\bibfnamefont {C.}~\bibnamefont {Punwong}}, \ and\ \bibinfo {author} {\bibfnamefont {T.~J.}\ \bibnamefont {Martínez}},\ }\href {\doibase https://doi.org/10.1016/j.chemphys.2010.03.020} {\bibfield  {journal} {\bibinfo  {journal} {Chemical Physics}\ }\textbf {\bibinfo {volume} {370}},\ \bibinfo {pages} {70} (\bibinfo {year} {2010})},\ \bibinfo {note} {dynamics of molecular systems: From quantum to classical}\BibitemShut {NoStop}%
\bibitem [{\citenamefont {Richings}\ and\ \citenamefont {Habershon}(2022)}]{richings2022}%
  \BibitemOpen
  \bibfield  {author} {\bibinfo {author} {\bibfnamefont {G.~W.}\ \bibnamefont {Richings}}\ and\ \bibinfo {author} {\bibfnamefont {S.}~\bibnamefont {Habershon}},\ }\href {\doibase 10.1021/acs.accounts.1c00665} {\bibfield  {journal} {\bibinfo  {journal} {Accounts of Chemical Research}\ }\textbf {\bibinfo {volume} {55}},\ \bibinfo {pages} {209} (\bibinfo {year} {2022})}\BibitemShut {NoStop}%
\bibitem [{\citenamefont {Worth}\ and\ \citenamefont {Cederbaum}(2004)}]{worth2004}%
  \BibitemOpen
  \bibfield  {author} {\bibinfo {author} {\bibfnamefont {G.}~\bibnamefont {Worth}}\ and\ \bibinfo {author} {\bibfnamefont {L.}~\bibnamefont {Cederbaum}},\ }\href {\doibase 10.1146/annurev.physchem.55.091602.094335} {\bibfield  {journal} {\bibinfo  {journal} {Annual Review of Physical Chemistry}\ }\textbf {\bibinfo {volume} {55}},\ \bibinfo {pages} {127} (\bibinfo {year} {2004})}\BibitemShut {NoStop}%
\bibitem [{\citenamefont {Liu}\ \emph {et~al.}(2020)\citenamefont {Liu}, \citenamefont {Horton}, \citenamefont {Yang}, \citenamefont {Nunes}, \citenamefont {Shen}, \citenamefont {Wolfe}, \citenamefont {Forbes}, \citenamefont {Cheng}, \citenamefont {Moore}, \citenamefont {Centurion}, \citenamefont {Hegazy}, \citenamefont {Li}, \citenamefont {Lin}, \citenamefont {Stolow}, \citenamefont {Hockett}, \citenamefont {Rozgonyi}, \citenamefont {Marquetand}, \citenamefont {Wang},\ and\ \citenamefont {Weinacht}}]{WOS:000527523000001}%
  \BibitemOpen
  \bibfield  {author} {\bibinfo {author} {\bibfnamefont {Y.}~\bibnamefont {Liu}}, \bibinfo {author} {\bibfnamefont {S.~L.}\ \bibnamefont {Horton}}, \bibinfo {author} {\bibfnamefont {J.}~\bibnamefont {Yang}}, \bibinfo {author} {\bibfnamefont {J.~P.~F.}\ \bibnamefont {Nunes}}, \bibinfo {author} {\bibfnamefont {X.}~\bibnamefont {Shen}}, \bibinfo {author} {\bibfnamefont {T.~J.~A.}\ \bibnamefont {Wolfe}}, \bibinfo {author} {\bibfnamefont {R.}~\bibnamefont {Forbes}}, \bibinfo {author} {\bibfnamefont {C.}~\bibnamefont {Cheng}}, \bibinfo {author} {\bibfnamefont {B.}~\bibnamefont {Moore}}, \bibinfo {author} {\bibfnamefont {M.}~\bibnamefont {Centurion}}, \bibinfo {author} {\bibfnamefont {K.}~\bibnamefont {Hegazy}}, \bibinfo {author} {\bibfnamefont {R.}~\bibnamefont {Li}}, \bibinfo {author} {\bibfnamefont {M.-F.}\ \bibnamefont {Lin}}, \bibinfo {author} {\bibfnamefont {A.}~\bibnamefont {Stolow}}, \bibinfo {author} {\bibfnamefont {P.}~\bibnamefont {Hockett}}, \bibinfo {author} {\bibfnamefont {T.}~\bibnamefont {Rozgonyi}},
  \bibinfo {author} {\bibfnamefont {P.}~\bibnamefont {Marquetand}}, \bibinfo {author} {\bibfnamefont {X.}~\bibnamefont {Wang}}, \ and\ \bibinfo {author} {\bibfnamefont {T.}~\bibnamefont {Weinacht}},\ }\href {\doibase 10.1103/PhysRevX.10.021016} {\bibfield  {journal} {\bibinfo  {journal} {Physical Review X}\ }\textbf {\bibinfo {volume} {10}} (\bibinfo {year} {2020}),\ 10.1103/PhysRevX.10.021016}\BibitemShut {NoStop}%
\end{thebibliography}%

\end{document}